\documentclass[a4paper,fleqn,usenatbib]{mnras}
\usepackage{longtable}
\usepackage{newtxtext,newtxmath}
\usepackage[T1]{fontenc}
\usepackage{ae,aecompl}
\usepackage{graphicx}	
\usepackage{amsmath}	
\usepackage{amssymb}	
\defcitealias{barker2018}{BP18}
\defcitealias{brasseur2010}{BSV10}
\defcitealias{ccm89}{CCM89}
\defcitealias{grundahl1998}{GVA98}
\defcitealias{ngc5904}{Paper I}
\defcitealias{paltrinieri1998}{PFF98}
\defcitealias{savino2018}{SMB18}
\defcitealias{sfd}{SFD}
\defcitealias{stetson2019}{SPZ19}

\title[Isochrone fitting of NGC\,6205]
{Isochrone fitting of Galactic globular clusters -- II. NGC\,6205 (M13)}

\author[G. A. Gontcharov, M. Yu. Khovritchev and A. V. Mosenkov]{
George A. Gontcharov,$^{1,2}$\thanks{E-mail: george.gontcharov@tdt.edu.vn}
Maxim Yu. Khovritchev$^{3,4}$
and Aleksandr V. Mosenkov$^{3}$
\\
$^{1}$Department for Management of Science and Technology Development, Ton Duc Thang University, Ho Chi Minh City, Vietnam\\
$^{2}$Faculty of Applied Sciences, Ton Duc Thang University, Ho Chi Minh City, Vietnam\\
$^{3}$Central Astronomical Observatory, Russian Academy of Sciences, 65/1 Pulkovskoye chaussee, St. Petersburg 196140, Russia\\
$^{4}$St. Petersburg State University, 7/9 Universitetskaya nab., St. Petersburg 199034, Russia
}

\date{Accepted 2020 June 10. Received 2020 June 9; in original form 2020 March 16}

\pubyear{2020}

\begin{document}
\label{firstpage}
\pagerange{\pageref{firstpage}--\pageref{lastpage}}
\maketitle

\begin{abstract}
We present new isochrone fits to colour--magnitude diagrams of the Galactic globular cluster NGC\,6205 (M13).
We utilise 34 photometric bands from the ultraviolet to mid-infrared by use of data from the 
{\it HST}, {\it Gaia} DR2, SDSS, unWISE, Pan-STARRS DR1, and other photometric sources.
In our isochrone fitting we use the PARSEC, MIST, DSEP, BaSTI, and IAC-BaSTI theoretical models and isochrones, 
both for the solar-scaled and He--$\alpha$--enhanced abundances, with a metallicity of about [Fe/H]$=-1.58$ adopted from 
the literature.
The colour--magnitude diagrams, obtained with pairs of filters from different datasets but of similar effective wavelengths,
show some colour offsets up to 0.04 mag between the fiducial sequences and isochrones.
We attribute these offsets to systematic differences of the datasets.
Some intrinsic systematic differences of the models/isochrones remain in our results:
the derived distances and ages are different for the ultraviolet, optical and infrared photometry used, while
the derived ages are different for the different models/isochrones, e.g.
in the optical range from $12.3\pm0.7$ Gyr for He--$\alpha$--enhanced DSEP to $14.4\pm0.7$ Gyr for MIST.
Despite the presence of multiple stellar populations, we obtain convergent estimates for the dominant population: 
best-fitting distance $7.4\pm0.2$ kpc, true distance modulus $14.35\pm0.06$ mag, 
parallax $0.135\pm0.004$ mas, extinction $A_\mathrm{V}=0.12\pm0.02$, and reddening $E(B-V)=0.04\pm0.01$.
These estimates agree with other recent estimates, however, the extinction and reddening are twice as high as generally 
accepted.
The derived empirical extinction law agrees with the Cardelli--Clayton--Mathis extinction law with the best-fitting 
$R_\mathrm{V}=3.1^{+1.6}_{-1.1}$.
\end{abstract}

\begin{keywords}
Hertzsprung--Russell and colour--magnitude diagrams --
dust, extinction --
globular clusters: general --
globular clusters: individual: NGC\,6205 (M13)
\end{keywords}

\section{Introduction}
\label{intro}

Multiple stellar populations have been found in all or almost all Galactic globular clusters (GCs) \citep{monelli2013}.
In particular, such populations can be seen using appropriate filters, colours, and colour combinations, which are sensitive to them. 
This makes it difficult to obtain accurate cluster properties from isochrone fits to colour--magnitude diagrams (CMDs).
Using a CMD, one must either indicate that a difference between the positions of the populations in that CMD is negligible or
that a dominant population exists and can be well separated in that CMD from other populations.
If this is the case for several CMDs, an age, distance, and reddening for such a population can be derived by isochrone fitting 
simultaneously in many ultraviolet (UV), optical and infrared (IR) bands for different stages of stellar evolution, 
which are well recognized in the CMDs, such as the main sequence (MS), its turn-off (TO), the subgiant branch (SGB), 
red giant branch (RGB), horizontal branch (HB), and asymptotic giant branch (AGB).
Moreover, such a fitting allows one to verify theoretical stellar evolution models underlying these isochrones.

Recently, we applied such a fitting to several CMDs of the Galactic GC NGC\,5904 \citep[][hereafter Paper I]{ngc5904}.
A dominant population in that cluster is evident and well separated in these CMDs.
We used the accurate photometry of individual stars of NGC\,5904 from the {\it Hubble Space Telescope (HST)}, 
{\it Gaia} DR2 \citep{gaiabrown},
{\it Wide-field Infrared Survey Explorer (WISE}, \citealt{wise}),
Sloan Digital Sky Survey (SDSS, \citealt{an2008}), 
Panoramic Survey Telescope and Rapid Response System (Pan-STARRS, \citealt{bernard2014}),
and other projects in 29 bands from the UV to mid-IR.
To fit the photometric data we used 
the PAdova and TRieste Stellar Evolution Code (PARSEC, \citealt{bressan2012}) \footnote{\url{http://stev.oapd.inaf.it/cgi-bin/cmd}},
the MESA Isochrones and Stellar Tracks (MIST, \citealt{paxton2011, paxton2013, mist, choi2016}) \footnote{\url{http://waps.cfa.harvard.edu/MIST/}},
the Dartmouth Stellar Evolution Program (DSEP, \citealt{dotter2007, dotter2008}) \footnote{\url{http://stellar.dartmouth.edu/models/}},
the Bag of Stellar Tracks and Isochrones (BaSTI, \citealt{basti2004, basti2006, basti2013}
\footnote{\url{http://albione.oa-teramo.inaf.it}}, and the IAC-BaSTI \citep{newbasti}
\footnote{\url{http://basti-iac.oa-abruzzo.inaf.it/index.html}}, as well as the isochrones from \citet{an2009}.
Adopting a spectroscopic metallicity value for NGC\,5904 from the literature, we derived the distance, age, and extinction law 
(a dependence of extinction on wavelength) for that cluster.

\citetalias{ngc5904} generally followed the approach of \citet{hendricks2012} and \citet[][hereafter BP18]{barker2018}, 
but we used many more bands to draw the extinction law and more models to verify the robustness of our results.
\citetalias{ngc5904} revealed that all the data and models for NGC\,5904, except for some UV and SDSS data, agree with the extinction law of 
\citet[][hereafter CCM89]{ccm89} with the extinction-to-reddening ratio $A_\mathrm{V}/E(B-V)\equiv R_\mathrm{V}=3.60\pm0.05$ and 
$A_\mathrm{V}=0.20\pm0.02$ mag.
This extinction is twice as high as generally accepted for that GC due to a rather high extinction between 625 and 2000 nm, i.e. 
between the optical and IR bands.
Such a high $R_\mathrm{V}$ had not been supposed for the Galactic GCs before \citet{hendricks2012} who first used IR data in their 
isochrone-to-data fit for the GC M4 and found a similar, rather high $R_\mathrm{V}=3.62\pm0.07$ for that cluster.

However, the large deviations of the isochrones from the data, found in \citetalias{ngc5904} for NGC\,5904,
can be explained not only by the high extinction and $R_\mathrm{V}$, but also by an intrinsic offset of the model colours due to an 
imperfection of the models, colour--$T_\mathrm{eff}$ relations and bolometric corrections used.
Hence, the approach and conclusions of \citetalias{ngc5904} should be verified for other GCs.

The aim of this study is to fit a multiband photometry and different isochrones in the CMDs for the GC NGC\,6205 (M13),
despite the presence of multiple populations in this cluster \citep{monelli2013, savino2018}.
Since \citetalias{barker2018} have investigated NGC\,6205 and used partially the same photometric data 
({\it HST} ACS/WFC3), we should use some additional data, especially in the IR, and pay more attention to agreement 
of the photometry from different sources.
Following \citetalias{ngc5904}, we adopt the spectroscopic metallicity and He--$\alpha$--enhancement from the literature 
in order to derive the most probable age, distance and empirical extinction law of NGC\,6205 by use of the photometry 
of its dominant population. Also, we estimate the accuracy of the isochrones under consideration.

This paper is organized as follows. We describe some key properties of NGC\,6205, including metallicity, in Sect.~\ref{metal}.
In Sect.~\ref{photo} we describe the photometry used, cleaning of the datasets and creation of the fiducial sequences in the CMDs.
In Sect.~\ref{iso} we describe the theoretical models and consequent isochrones used.
Some results of our isochrone fitting are presented in Sect.~\ref{results} with an adjustment of the colours in Sect.~\ref{adjustment}.
We summarize our main findings and conclusions in Sect.~\ref{conclusions}.
We present an analysis of the uncertainties in Appendix~\ref{uncertainties} and some additional CMDs in Appendix~\ref{cmds}.

\section{Some properties of NGC\,6205}
\label{metal}

NGC 6205 (M13) is one of the brightest and most well-studied GCs of our Galaxy, located in the constellation of Hercules,
at RA(2000)$=16^h41^m41^s$ and DEC(2000)$=+36\degr27\arcmin36\arcsec$ \citep{goldsbury2010, miocchi2013} or 
$l=59.0073\degr$ and $b=+40.9131\degr$.

We select NGC\,6205 for this study due to its rich multiband photometry, low foreground and differential reddening, accurate spectroscopic 
estimates of its metallicity, and previous successful isochrone fitting of its dominant population, e.g. by \citetalias{barker2018}.

\citet{bonatto2013} obtained the differential reddening for this GC $\delta E(B-V)=0.033\pm0.09$ and $\max\delta E(B-V)=0.068$.
Their note that values lower than $\delta E(B-V)<0.04$ in their analysis of many GCs may be due to photometric zero-point variations 
is apparently related to NGC\,6205.
However, \citet{bonatto2013} studied only part of the cluster's field within about 3.3 arcmin from its centre, which is covered by 
the {\it HST} photometry used. Recently, Bonatto (private communication) analysed NGC\,6205's photometry from \citet[][hereafter SPZ19]{stetson2019} 
and revealed that the northern and north-western parts of a field within 16 arcmin from the cluster's centre show few hundredths 
magnitude higher $E(B-V)$ than the southern and south-eastern parts. This will be discussed in Sect.\ref{photo}.

The commonly used database of GCs by \citet{harris}\footnote{\url{https://www.physics.mcmaster.ca/~harris/mwgc.dat}}, 2010 revision,
provides for NGC\,6205 a distance of 7.64 kpc, reddening $E(B-V)=0.02$\,mag, and apparent visual distance modulus $(m-M)_\mathrm{V}=14.42$.
However, both distance and age estimates from the literature show a considerable diversity 
\citepalias[][and references therein]{barker2018}.

The compiled estimate $E(B-V)=0.02$\,mag by \citet{harris} is based on some original estimates, such as $E(b-y)=0.015\pm0.010$, 
i.e. $E(B-V)=0.021\pm0.014$ mag by \citet{grundahl1998}, which is compared with our results in Sect.~\ref{extlaw}.

A recent Bayesian single-population analysis of the {\it HST} photometry in the $F606W$ and $F814W$ bands by \citet{wagner-kaiser2017} gives 
[Fe/H]$=-1.53$, $(m-M)_\mathrm{V}=14.442^{+0.006}_{-0.005}$, an age of $13.094^{+0.084}_{-0.082}$,
$A_\mathrm{V}=0.12^{+0.005}_{-0.006}$, and, consequently, $E(B-V)=0.039\pm0.002$ mag by use of the extinction law of \citetalias{ccm89} with 
$R_\mathrm{V}=3.1$.
Moreover, analysing 69 GCs, \citet{wagner-kaiser2017} note, `In many cases, we estimate moderately larger absorption than the values 
from Harris (2010)'.

A recent isochrone fitting of $VI$ photometry by \citet{deras2019} gives $E(B-V)$, age and distance of 0.02 mag, 12.6 Gyr and 7.1 kpc, 
respectively.
However, \citet{deras2019} note that the HB needs $E(B-V)=0.04$ mag for a better fit.

We adopt the spectroscopic estimate of metallicity [Fe/H]$=-1.58\pm0.04$ (equal to $Z=0.0004$ with the solar metallicity $Z=0.0152$) 
derived for NGC\,6205 by \citet{carretta2009} within their abundance scale, which is well defined from a robust dataset of [Fe/H] 
abundances of GCs.
This estimate agrees with recent estimates from a Fourier decomposition of the light curves of RRab and RRc stars by \citet{deras2019}, 
[Fe/H]$=-1.58\pm0.09$, and from a review of chemical composition of Galactic GCs by \citet{marsakov2019}, [Fe/H]$=-1.54\pm0.06$.

The estimate of metallicity from spectroscopy is much more accurate than that from photometry. 
Hence, we do not determine metallicity from our isochrone fitting in the CMDs.
However, we check a range of metallicity, outside of which the derived cluster's characteristics and the isochrone fitting, 
in particular, the tilt of the RGB, become completely unreliable; this range is $-1.38$<[Fe/H]$<-1.78$.
Thus, the results of this study do not contradict the adopted metallicity.

\cite{marino2019, marsakov2019} find an enhanced average $\alpha$ abundance of [$\alpha/$Fe]$\approx0.2$ for NGC\,6205, with
noticeable variations from star to star and from element to element.
\citet{denissenkov2015, cohen2017, denissenkov2017, wagner-kaiser2017} show a significant ($0.30<Y<0.38$) helium enhancement in NGC\,6205.
\citet{denissenkov2015} note three populations of its stars: with normal ($Y=0.25$), intermediate ($Y=0.33$), and extreme ($Y=0.38$) 
compositions, with a dominance of the intermediate one with $>60$ per cent of the stars.
Based on Str\"omgren photometry and its comparison with several spectroscopic studies, \citet[][hereafter SMB18]{savino2018} 
conclude that about 80 per cent of 
giant stars of NGC\,6205 belong to the He--$\alpha$--enriched population.
Such a clear domination of the population makes the photometry of this cluster quite convenient for an isochrone fitting.

Since $\alpha$ and helium enhancements make isochrones with the same [Fe/H] redder and bluer, respectively 
(see \citetalias{ngc5904} for details), a cumulative effect of both the enhancements may keep the isochrones close to 
the solar-scaled ones.
In Sect.~\ref{results} we will test the difference between the solar-scaled and He--$\alpha$--enhanced isochrones and, 
consequently, the importance of taking into account the He--$\alpha$--enhancement.
The contribution of the uncertainty due to the adopted He--$\alpha$--enhancement to the total uncertainty is discussed in 
Appendix~\ref{uncertainties}.

\section{Photometry}
\label{photo}

We use the following datasets of photometry in 34 filters in order to create CMDs for NGC\,6205:
\begin{itemize}
\item 54\,814 stars
\footnote{The original datasets typically contain many more stars. Here we provide the number of stars after rejecting the poor 
photometry, as described in Sect.~\ref{cleaning}.}
with photometry in the $F275W$, $F336W$, $F438W$ filters from the {\it HST} Wide Field Camera 3 (WFC3) UV Legacy Survey of Galactic 
Globular Clusters and the $F606W$, $F814W$ filters from its Advanced Camera for Surveys (ACS) survey of Galactic globular clusters
\citep{nardiello2018}\footnote{\url{http://groups.dfa.unipd.it/ESPG/treasury.php}},
\item 25\,032 stars with the SDSS photometry in the $ugriz$ filters and related fiducial sequences from \citet{an2008, an2009}
\footnote{\url{http://classic.sdss.org/dr6/products/value_added/anjohnson08_clusterphotometry.htm}
Their fiducial sequences do not expand on the HB. Therefore, we calculate our own fiducial sequences for the HB by use of these data,
however, for the RGB, SGB, TO and MS our fiducial sequences appear to coincide with those of \citet{an2008,an2009} within $\pm0.01$ mag.},
\item 23\,023 stars with $UBVI$ photometry \citepalias{stetson2019} \footnote{\url{http://cdsarc.u-strasbg.fr/viz-bin/cat/J/MNRAS/485/3042}
We do not use the photometry in the $R$ filter due to its very low precision.},
\item 18\,685 stars with Str\"omgren $vy$ photometry from the Isaac Newton Telescope -- Wide Field Camera (INT/WFC) \citepalias{savino2018}
\footnote{We do not use the photometry from \citetalias{savino2018} in the Str\"omgren $b$ band
due to a problem with the data, which is noted by the authors.},
\item 12\,868 stars with Str\"omgren $uvby$ photometry from the Nordic Optical Telescope (NOT) on La Palma, Canary Islands 
\citep[][hereafter GVA98]{grundahl1998},
\item 6747 stars with photometry from \citet{piotto2002} in the $F439W$ and $F555W$ filters from the {\it HST} Wide Field and 
Planetary Camera 2 (WFPC2),
\item 4530 stars with {\it Gaia} DR2 photometry in the $G$, $G_\mathrm{BP}$ and $G_\mathrm{RP}$ filters from \citet{bustos2019}
\footnote{\url{http://cdsarc.u-strasbg.fr/viz-bin/cat/J/MNRAS/488/3024}},
\item 2152 stars and fiducial sequences derived by \citet{rey2001} in the $BV$ bands, based on the Michigan--Dartmouth--MIT (MDM) 
Observatory 2.4 m telescope \footnote{\url{http://cdsarc.u-strasbg.fr/viz-bin/cat/J/AJ/122/3219}},
\item 1699 stars with {\it Wide-field Infrared Survey Explorer (WISE)} photometry in the $W1$ filter from unWISE catalogue \citep{unwise}, 
which are cross-identified by us with the other lists under consideration,
\item the fiducial sequences derived by \citet{clem2008} in the $u'g'r'i'z'$ bands with the MegaCam wide-field imager on the 
Canada--France--Hawaii Telescope (CFHT/MegaCam),
\item the fiducial sequence derived by \citet[][hereafter PFF98]{paltrinieri1998} for the $B-V$ colour
based on observations with the 1.23-m telescope at the German--Spanish Astronomical Center, Calar Alto, Spain (hereafter GSAC 1.23-m),
\item the fiducial sequences derived by \citet[][hereafter BSV10]{brasseur2010} for the $V-J$ and $V-K$ colours based 
on IR observations with the WIRCam imager on the Canada--France--Hawaii Telescope (CFHT/WIRCam), 
calibrated by use of the Two Micron All-Sky Survey (2MASS, \citealt{2mass}) stars, and 
combined with an early version of the $V$-band data from \citetalias{stetson2019},
\item the fiducial sequences derived by \citet{bernard2014} in the $g_\mathrm{P1}$, $r_\mathrm{P1}$, $i_\mathrm{P1}$, $z_\mathrm{P1}$, and $y_\mathrm{P1}$ 
bands of Pan-STARRS DR1.
\end{itemize}

Table~\ref{filters} presents the effective wavelength $\lambda_\mathrm{eff}$ in nm and the median precision of the photometry for each filter.

\begin{table}
\def\baselinestretch{1}\normalsize\small
\caption[]{The effective wavelength (nm) and median precision of the photometry (mag) for the filters under consideration.
}
\label{filters}
\[
\begin{tabular}{llcc}
\hline
\noalign{\smallskip}
 Telescope & Filter & $\lambda_\mathrm{eff}$ & Median precision \\
\hline
\noalign{\smallskip}
{\it HST}/WFC3           & $F275W$                & 274 & 0.02 \\
{\it HST}/WFC3           & $F336W$                & 329 & 0.01 \\
INT/WFC \citepalias{savino2018}    & Str\"omgren $u$        & 347 & 0.02 \\ 
NOT \citepalias{grundahl1998} & Str\"omgren $u$        & 347 & 0.02 \\
Various \citepalias{stetson2019} & $U$            & 354 & 0.01 \\
SDSS                     & $u$                    & 360 & 0.06 \\
INT/WFC \citepalias{savino2018}    & Str\"omgren $v$        & 412 & 0.01 \\ 
NOT \citepalias{grundahl1998} & Str\"omgren $v$        & 412 & 0.01 \\
{\it HST}/WFPC2          & $F439W$                & 435 & 0.03 \\
Various \citepalias{stetson2019} & $B$            & 437 & 0.01 \\
MDM \citep{rey2001}      & $B$                    & 437 & 0.01 \\
GSAC 1.23-m \citepalias{paltrinieri1998} & $B$         & 437 & 0.05 \\
{\it HST}/WFC3           & $F438W$                & 437 & 0.01 \\
NOT \citepalias{grundahl1998} & Str\"omgren $b$        & 467 & 0.01 \\
SDSS                     & $g$                    & 471 & 0.02 \\
Pan-STARRS1              & $g_\mathrm{P1}$        & 480 & 0.02 \\
CFHT/MegaCam             & $g'$                   & 491 & 0.02 \\
{\it Gaia} DR2           & $G_\mathrm{BP}$        & 539 & 0.03 \\
Various \citepalias{stetson2019} & $V$            & 547 & 0.01 \\
MDM \citep{rey2001}      & $V$                    & 547 & 0.01 \\
GSAC 1.23-m \citepalias{paltrinieri1998} & $V$         & 547 & 0.05 \\
INT/WFC \citepalias{savino2018}    & Str\"omgren $y$        & 548 & 0.02 \\ 
NOT \citepalias{grundahl1998} & Str\"omgren $y$        & 548 & 0.02 \\
{\it HST}/WFPC2          & $F555W$                & 549 & 0.08 \\
{\it HST}/ACS            & $F606W$                & 588 & 0.01 \\
Pan-STARRS1              & $r_\mathrm{P1}$        & 620 & 0.02 \\
SDSS                     & $r$                    & 621 & 0.03 \\
CFHT/MegaCam             & $r'$                   & 625 & 0.02 \\
{\it Gaia} DR2           & $G$                    & 642 & 0.02 \\
SDSS                     & $i$                    & 743 & 0.04 \\
Pan-STARRS1              & $i_\mathrm{P1}$        & 746 & 0.02 \\
{\it Gaia} DR2           & $G_\mathrm{RP}$        & 767 & 0.03 \\
CFHT/MegaCam             & $i'$                   & 767 & 0.02 \\
{\it HST}/ACS            & $F814W$                & 794 & 0.01 \\
Various \citepalias{stetson2019} & $I$            & 812 & 0.01 \\
Pan-STARRS1              & $z_\mathrm{P1}$        & 860 & 0.02 \\
CFHT/MegaCam             & $z'$                   & 885 & 0.02 \\
SDSS                     & $z$                    & 885 & 0.06 \\
CFHT/WIRCam, 2MASS       & $J$                    & 1229 & 0.02 \\
CFHT/WIRCam, 2MASS       & $K$                    & 2147 & 0.02 \\
{\it WISE}               & $W1$                   & 3316 & 0.01 \\

\hline
\end{tabular}
\]
\end{table}


\begin{figure}
\includegraphics{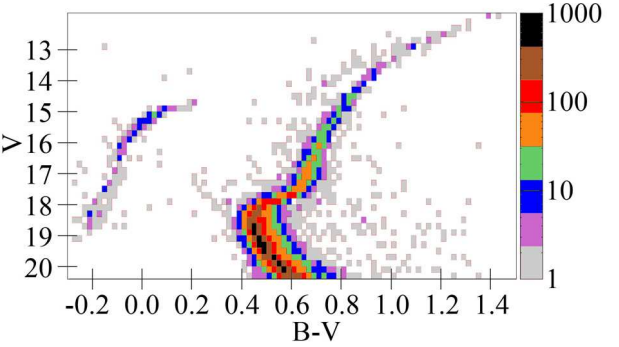}
\caption{The distribution of 23\,023 stars from \citetalias{stetson2019} in the colour--magnitude bins of $0.02\times0.2$ mag on the
$B-V$ versus $V$ CMD of NGC\,6205.
The number of the stars in each bin is shown by the colour scale on the right.
}
\label{bvv}
\end{figure}

Each star has a photometry in some but not all filters.

We cross-identify some of the datasets with each other. The cross-identification in the UV and optical ranges
is done to estimate possible offsets and systematic errors of the photometry of the different datasets,
as discussed in Sect.~\ref{adjustment} (as a result, we decide to use the Str\"omgren datasets by \citetalias{savino2018}
and \citetalias{grundahl1998} together due to their negligible photometric differences for common stars).
The cross-identification with unWISE is done to determine some IR extinction zero-points in Sect.~\ref{extlaw} by use of the 
optical--IR colours and reddenings, following the approach of \citetalias{ngc5904}.
 
The following datasets have an IR extinction zero-point:
$UBVI$ by \citetalias{stetson2019} based on the derived reddening $E(I-W1)$ or $E(V-W1)$,
$BV$ by \citet{rey2001} based on the derived $E(V-W1)$,
{\it Gaia} based on the derived $E(G-W1)$ or $E(G_\mathrm{RP}-W1)$,
SDSS based on the derived $E(z-W1)$,
Str\"omgren by \citetalias{savino2018} and \citetalias{grundahl1998} based on the derived $E(y-W1)$, and 
$VJK$ by \citetalias{brasseur2010} based on the derived $E(V-K)$.
The remaining datasets, including all the cases when we have fiducial sequences only, without data for individual stars,
have not been tied to an IR photometric catalogue:
{\it HST} ACS/WFC3, {\it HST} WFPC2, $BV$ by \citetalias{paltrinieri1998}, Pan-STARRS by \citet{bernard2014}, 
and CFHT/MegaCam by \citet{clem2008}.
Their extinction zero-points are defined by the \citetalias{ccm89} extinction law with $R_\mathrm{V}$ determined 
from the datasets with an IR extinction zero-point, as described in Sect.~\ref{extlaw}.

These vast amount of photometric data make it possible to consider and fit the isochrones to dozens of CMDs with different colours.
They cover a wide wavelength range between the UV and middle IR.
Each CMD provides us with independent estimates of age and distance, while each dataset gives us an irrespective 
set of the derived reddenings.
Such a set draws an empirical extinction law for a dataset with an IR extinction zero-point, while each dataset 
without such a zero-point can be used as a set of reddenings to confirm or reject this law.

To fit by isochrones, a fiducial sequence must be determined in each CMD as a colour--magnitude relation for single 
stars of a dominant population in a cluster.
We calculate the fiducial sequence for each CMD as a locus of the number density maxima in some colour--magnitude bins.

An example of the distribution of the stars in a CMD, after the cleaning for poor photometry described in 
Sect.~\ref{cleaning}, is shown in Fig.~\ref{bvv}: 
$B-V$ versus $V$ CMD with the colour--magnitude bins of $0.02\times0.2$ mag.
This is an optimal bin size for a rather dense distribution. 
It is evident from Fig.~\ref{bvv} that in such a case a fiducial sequence can be derived, at least, with a precision 
of half of one bin.
For a less dense distribution (for example, in the CMDs with the $W1$ band) the bin size increases to $0.04\times0.4$ mag.
Consequently, it is evident that an average fiducial sequence colour (averaged over $>15$ CMD magnitude bins
and finally transformed into the reddening in a comparison with an isochrone)
can be determined with a precision better than $0.04/2/15^{0.5}\approx0.005$ mag. 
An example fiducial sequence for $V$ versus $B-V$ is presented in Table~\ref{fiducial}. 
All other fiducial sequences can be provided on request.

\begin{figure}
\includegraphics{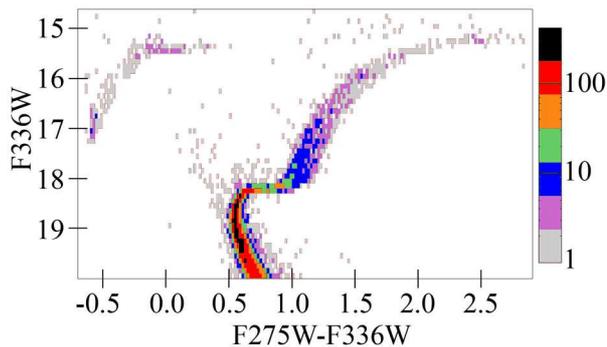}
\caption{The distribution of 24\,303 stars with the $F275W$ and $F336W$ {\it HST} WFC3 photometry in the colour--magnitude bins of
$0.02\times0.1$ mag on the $F275W-F336W$ versus $F336W$ CMD of NGC\,6205.
The number of stars in each bin is shown by the colour scale on the right.
}
\label{f275_336}
\end{figure}

\begin{table}
\def\baselinestretch{1}\normalsize\normalsize
\caption[]{The fiducial sequence for NGC\,6205 $V$ versus $B-V$ based on the data of \citetalias{stetson2019}.
The complete table is available online.
}
\label{fiducial}
\[
\begin{tabular}{rr}
\hline
\noalign{\smallskip}
$V$ & $B-V$ \\
\hline
\noalign{\smallskip}
18.6 & -0.210 \\
18.4 & -0.190 \\
18.2 & -0.170 \\
18.0 & -0.150 \\
17.7 & -0.130 \\
\ldots & \ldots \\
\hline
\end{tabular}
\]
\end{table}


Since we use a similar number of stars for all the filters of a dataset, and since we need a similar and rather large
number of stars in the bins of number density maxima to obtain a precise fiducial sequence position, 
the optimal bin size depends on the wavelength range used.
For example, if we consider a bin of 0.02 mag for both the $B-V$ and $V-I$ colours, we should use a bin size of 0.04 mag for the 
$B-I$ colour.

Similar to NGC\,5904 in \citetalias{ngc5904}, we can easily define the fiducial sequences for NGC\,6205 due to
(i) the high degree of completeness of the stellar samples under consideration, at least between the HB and TO,
(ii) low differential reddening,
(iii) low contamination from foreground/background stars at such a high latitude, and
(iv) the high percentage of the dominant population stars.

UV CMDs are difficult to use for isochrone fitting, particularly due to an emphasizing of the colour differences 
between the stellar populations by some combinations of UV filters \citep{monelli2013, barker2018, savino2018}.
However, it is interesting to explore with recent data and isochrones -- which combination of the UV filters makes the 
difference between the populations being negligible or allows us to derive robust properties of a dominant population.
Therefore, we use UV filters in our study as well.

A UV CMD, after cleaning for poor photometry in Sect.~\ref{cleaning}, is shown in Fig.~\ref{f275_336}.
It illustrates a separation of the multiple populations only for the RGB, making it wider.
This makes a different contribution to the uncertainties of the derived reddening, distance and age.

A dominant population with about 80 per cent of the stars, as predicted by \citetalias{savino2018}, formes a peak
of the stellar distribution along the bluer side of the RGB.
This peak has a width up to four colour bins (i.e. 0.08 mag) in the cross-sections of the RGB along the colour.
Hence, the fiducial sequence colour in such a cross-section (i.e. within a magnitude bin) can be determined with 
a precision of half of the peak width, i.e., at least, $\pm0.04$ mag.
To derive the reddening, a fiducial sequence colour, averaged over many magnitude bins, is used. 
For example, for the CMD in Fig.~\ref{f275_336} we use $40$ magnitude bins within $15.5<F336W<19.5$ mag.
In each magnitude bin we have an independent estimate of the fiducial colour.
Hence, a typical mean fiducial sequence colour uncertainty is $0.04/40^{0.5}=0.006$ mag.

To derive the distance, the magnitudes of the HB and SGB are used.
The CMD in Fig.~\ref{f275_336} is typical showing no influence of the multiple populations on the HB or SGB magnitude
and, consequently, on the distance.
The same is true for the age, when it is obtained from the HB--SGB magnitude difference.
However, when it is obtained from the SGB length, it can be affected by a separation of the multiple populations 
at the base of the RGB.
The CMD in Fig.~\ref{f275_336} shows an uncertainty of the SGB length about half of the colour bin, i.e. $\pm0.01$ mag.
This uncertainty is typical in the UV, decreasing in the optical and IR range.
This corresponds to an age uncertainty of $\pm0.5$ Gyr. It gives a minor contribution w.r.t. the other uncertainties.

Finally, we find a rather precise positioning of the dominant population in all the CMDs.
The largest uncertainty 0.01 mag for the fiducial sequence colour, due to the multiple populations, 
is found for the $u-v$ versus $v$ CMD with the dataset of \citetalias{grundahl1998}.

These uncertainties due to the existence of multiple populations in NGC\,6205 are taken into account in 
the balance of uncertainties, presented in Appendix~\ref{uncertainties}.

\subsection{Cleaning the datasets}
\label{cleaning}

\begin{figure}
\includegraphics{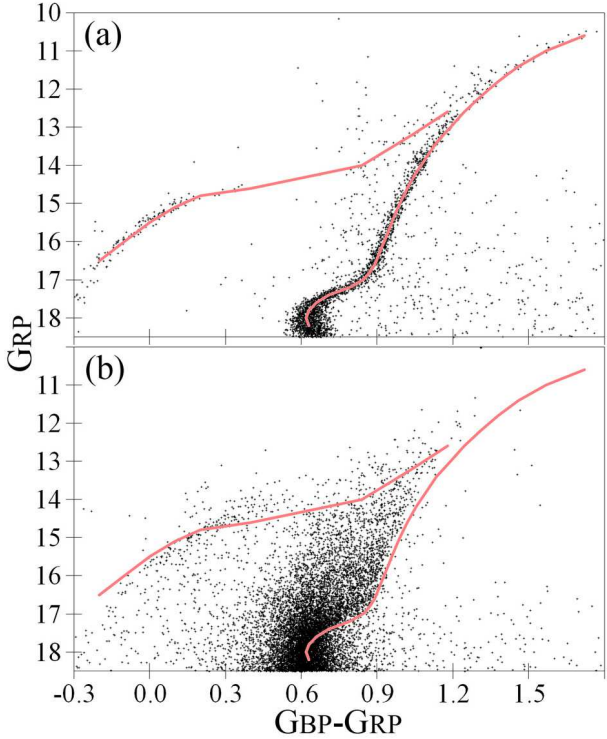}
\caption{$G_\mathrm{BP}-G_\mathrm{RP}$ versus $G_\mathrm{RP}$ CMD of NGC\,6205 for (a) 4844 and (b) 21\,805 stars, 
which follow the limit~(\ref{excess}) or do not, respectively.
The fiducial sequence is shown by the curve as a reference.
}
\label{gaia_good_bad}
\end{figure}

\begin{figure}
\includegraphics{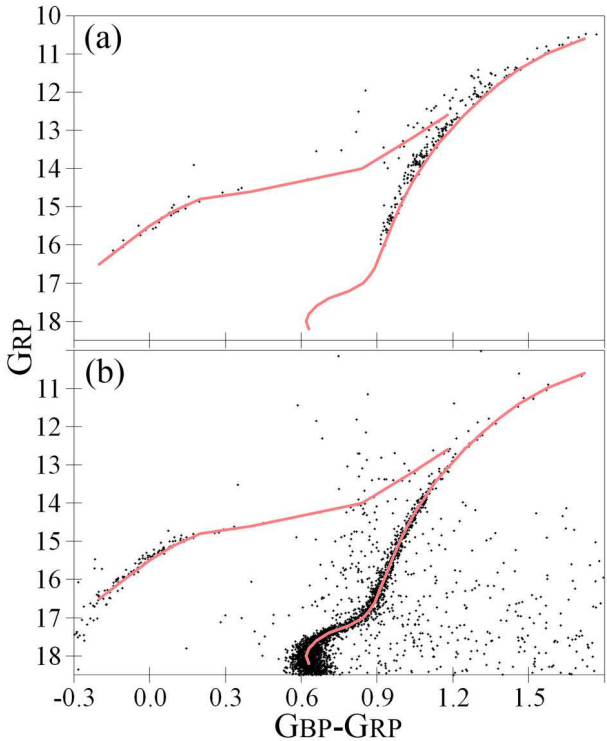}
\caption{The same as Fig.~\ref{gaia_good_bad} but for (a) 314 stars within and (b) 4530 stars outside the 2.7 arcmin 
radius from the cluster's centre.
The fiducial sequence is shown by the curve as a reference.
}
\label{gaia_radius}
\end{figure}

In this section we describe how we clean the original datasets from poor photometry.

We analyse the distribution of stars on the precision of their photometry and only select stars with errors of 
less than 0.07~mag in the {\it HST} ACS and WFC3 colours, 0.07~mag in \citet{rey2001}'s colour,
0.08~mag in the {\it Gaia} colour, 0.10~mag in the $UBVI$ colours, and 0.15~mag in the remaining colours.
However, the median precision of the colours is much higher, as is evident from Table~\ref{filters}.
Also, for each colour we only select stars brighter than a magnitude limit, in which 97 per cent of stars have a 
photometric error less than the above-mentioned limits. 
Typically, this means a magnitude limit of about $<20.5$ mag for the optical filters.
These magnitude cuts allow us to consider only the magnitude ranges where there is no bias of the fiducial sequence 
due to the removal of stars with poor photometry.

In addition, we use some photometric quality parameters to clean the datasets from poor photometry.
We present here only some interesting cases of such cleaning.

Initially we select 43\,714 {\it Gaia} DR2 stars in the field of NGC\,6205 by use of the catalogue of 
\citet{bustos2019}.
However, only 26\,649 stars among these have photometry in all three {\it Gaia} bands, which is needed to estimate 
the photometric quality.
Moreover, only 4844 stars among them follow the photometric quality criterion, suggesting that they are 
`well-behaved single sources' 
\citep{gaiaevans}:
\begin{eqnarray}
\label{excess}
&\verb"phot_bp_rp_excess_factor"\equiv \nonumber\\
&\equiv (F_\mathrm{BP}-F_\mathrm{RP})/F_\mathrm{G}<1.3+0.06(G_\mathrm{BP}-G_\mathrm{RP})^2\,,
\end{eqnarray}
where $F_\mathrm{BP}$, $F_\mathrm{RP}$, and $F_\mathrm{G}$ are the fluxes within the {\it Gaia} filters.

To justify the removal of such a large amount of stars, we compare the CMDs with the subsamples separated by the 
limit~(\ref{excess}) in Fig.~\ref{gaia_good_bad}.
It is seen that any relaxing of the limit~(\ref{excess}) would lead to a systematic blueward colour offset of the 
derived fiducial.
Thus, \verb"phot_bp_rp_excess_factor" appears to be a useful photometric quality parameter in order to separate 
stars with good or bad photometry in such studies of globular clusters.

However, the resulting dataset of the {\it Gaia} DR2 photometry still contains a number of stars with erroneous 
photometry in the most crowded centre of the cluster.
This is seen in Fig.~\ref{gaia_radius}, where 314 stars within 2.7 arcmin from the centre and the remaining 4530 stars 
within 16 arcmin from the centre are shown in plots (a) and (b), respectively: a systematic colour offset of the 
observed RGB by about 0.02 mag is evident in plot (a) w.r.t. (b).
These 314 stars dominate at brighter parts of the RGB and AGB, higher than a point where they meet each other, 
i.e. at $G_\mathrm{RP}<14$.
Therefore, we do not use these stars and these brighter parts of the RGB and AGB in calculating the fiducial 
sequence and in the fiducial--isochrone fitting.
This case indicates that the {\it Gaia} DR2 photometry may be unacceptable in central regions of GCs.

Note that such poor photometry is typical for brighter parts of the RGB and AGB of all the datasets under 
consideration. Therefore, we do not use these RGB and AGB parts in our isochrone fitting for all the datasets.

\begin{figure}
\includegraphics{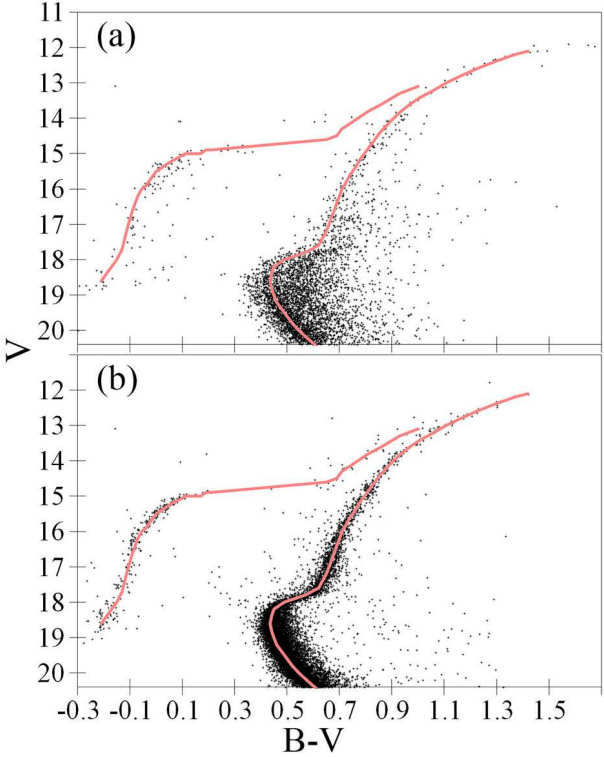}
\caption{$B-V$ versus $V$ CMD of NGC\,6205 based on the data from \citetalias{stetson2019} for (a) 4767 stars with 
variability evidence $>9$ and (b) 23\,023 remaining stars.
The fiducial sequence is shown by the curve as a reference.
}
\label{bv_init}
\end{figure}

\begin{figure}
\includegraphics{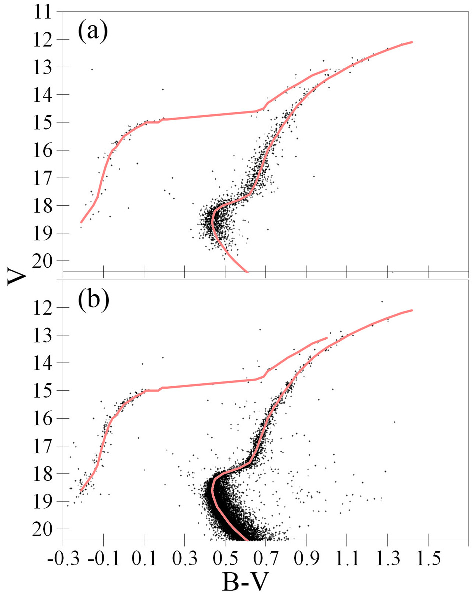}
\caption{The same as Fig.~\ref{bv_init} but for (a) 1852 stars within and (b) 21\,171 stars outside the 2.4 arcmin 
radius from the cluster's centre.
The fiducial sequence is shown by the curve as a reference.
}
\label{bv_radius}
\end{figure}

\begin{figure}
\includegraphics{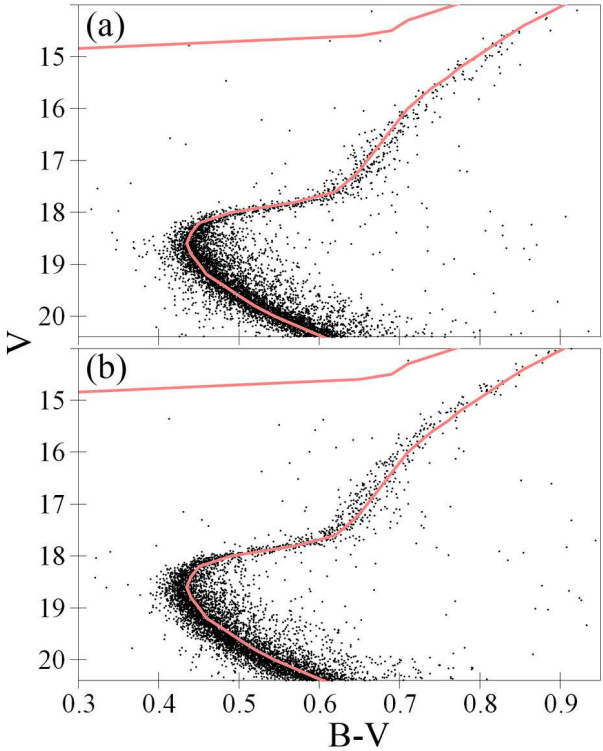}
\caption{The same as Fig.~\ref{bv_init} but for (a) 8130 stars in the north-west ($Y-X>164$ pixels in the 
\citetalias{stetson2019}'s coordinate frame) and (b) 7970 stars in the south-east ($Y-X<-167$ pixels in the 
\citetalias{stetson2019}'s coordinate frame) corner of NGC\,6205's field.
The fiducial sequence is shown by the curve as a reference.
}
\label{bv_dr}
\end{figure}

\begin{figure}
\includegraphics{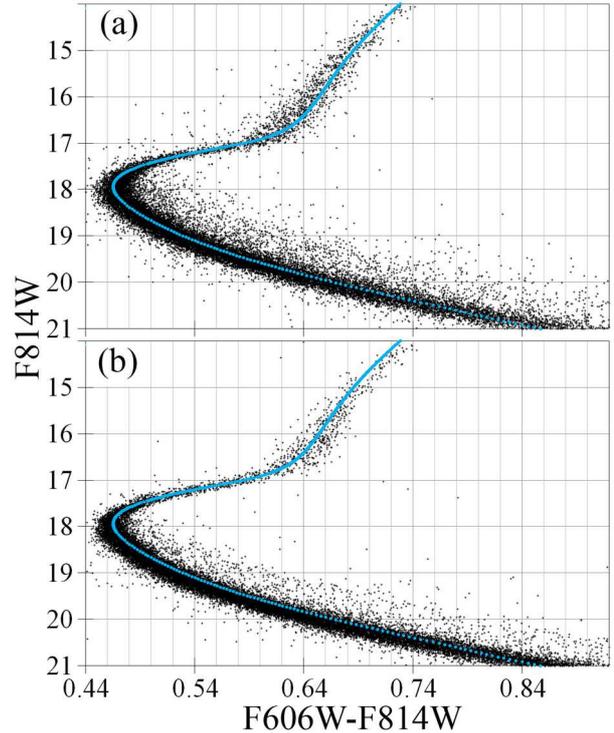}
\caption{$F606W-F814W$ versus $F814W$ CMD of NGC\,6205 based on the {\it HST} ACS data for (a) 25\,231 stars within and 
(b) 24\,958 stars outside the 1.14 arcmin radius from the cluster's centre. The best-fitting IAC-BaSTI isochrone is 
shown by the curve as a reference.
}
\label{hst_radius}
\end{figure}

For the dataset of \citetalias{stetson2019} there are several quality parameters to be taken into account to 
separate stars with accurate photometry.
In the field of NGC\,6205 we remove more than 40\,000 stars with DAOPHOT sharp parameter $|{\tt sharp}|>0.3$.
In addition, we remove 4767 stars with a variability evidence $>9$, which is defined by \citetalias{stetson2019} 
as the logarithm of the Welch--Stetson variability index \citep{welchstetson} times its weight.
These removed stars are shown in Fig.~\ref{bv_init} (a), while the remaining 23\,023 stars are in plot (b), 
with their precise distribution shown in Fig.~\ref{bvv}.
It is seen that the removed stars would provide a noticeable redward bias of the fiducial.

The resulting CMDs for the stars within and outside the 2.4 arcmin radius from the cluster's centre are shown in Fig.~\ref{bv_radius}.
Although a lot of faint stars are lost in the cluster's centre, the remaining stars demonstrate only a larger 
scatter around the fiducial sequence, while the systematic blueward colour offset of the remaining RGB stars in the centre w.r.t. the 
periphery is less than 0.01 mag.
Such a colour offset may be due to a lower reddening in the centre, in agreement with the results of Bonatto (private communication).
This shows that any centre--periphery distinction cannot be separated from the differential reddening.

Fig.~\ref{bv_dr} shows the differential reddening for the \citetalias{stetson2019} dataset within 25 arcmin from the cluster's centre.
It is seen that both the RGB and MS are about 0.02 mag redder in the north-western than in the south-eastern 
corner of the NGC\,6205's field.
This agrees with the differential reddening found by Bonatto (private communication) and mentioned in Sect.~\ref{intro}.
Therefore, it seems that the differential reddening dominates among the field effects after the cleaning of 
the \citetalias{stetson2019} dataset.
However, for other datasets and CMDs, such as the {\it HST} ACS/WFC3 dataset, a combination of the field effects 
may be more complex.

From the {\it HST} ACS/WFC3 dataset, we only use stars with $|{\tt sharp}|<0.15$, following \citet{nardiello2018}. 
This removes the majority of stars from the original dataset.
Also, we remove stars with a membership probability $<90$ per cent.
However, we use stars with the undefined probability, since they make up a significant portion of the sample and 
do not show a systematic offset in any CMD. Finally, we only use stars with a quality-fitting parameter $>0.9$.

However, the resulting sample of 54\,814 stars still demonstrates some systematic effects, even in the small 
{\it HST} field of $3.5\times3.5$ arcmin.
Fig.~\ref{hst_radius} shows the $F606W-F814W$ versus $F814W$ CMD for the stars within and outside the 1.14 arcmin 
radius from the cluster's centre in plots (a) and (b), respectively.
We show only a central part of the CMD, where the effects are noticeable (they are negligible at the HB and SGB). 
The best-fitting IAC-BaSTI isochrone is shown as a reference.
It is seen that in the centre of the cluster's field the RGB is bluer, while the MS is redder than in the 
periphery of the field.
This may be explained only by a combination of effects, such as a crowding effect, differential reddening, 
different representation of the multiple populations in different parts of the cluster, 
a systematic error of photometry over the field, a variable contamination by field stars, and/or an incompleteness 
of the sample.
The latter is especially noticeable for the faintest stars ($F814W>20$ mag) in Fig.~\ref{hst_radius} (a) w.r.t. (b), i.e. 
in the centre w.r.t. the periphery, apparently, due to a crowding effect.
Thus, these effects are difficult to separate and have to be considered mutually.

We briefly analyse these field effects for all the CMDs, 
which are based on the largest datasets of {\it HST} ACS/WFC3, SDSS, \citetalias{stetson2019}, 
\citetalias{grundahl1998}, and \citetalias{savino2018}. 
First, we compare partial CMDs, i.e. the ones for different parts of the field, by use of a moving window with 
about 5000 stars to calculate a partial fiducial sequence for each partial CMD.
The number of stars in the moving window varies slightly depending on their distribution over the field. 
We consider the shifts of the partial fiducial sequences w.r.t each other along the reddening vector as estimates of the
reddening and related field effects in different parts of the field.
This approach allows us to reveal only a gradient, but not a detailed map of the field effects. It seems to be enough,
since we find a rather small field effect at a level of $\delta E(B-V)<0.02$ mag for all the CMDs.
This agrees with the differential reddening results of \citet{bonatto2013} and Bonatto (pivate communication).
Given such a small effect for the largest datasets, we suggest a similar one for the remaining datasets.
Also, we check that stars from the areas with bluer and redder partial fiducial sequences are equally represented in all the 
CMDs for all the datasets, or, at least, that an uneven distribution of stars does not introduce a field effect bias more than $0.005$ mag 
into an average fiducial sequence colour. We find that, indeed, this is so for all the datasets.

We can conclude that a typical variation of the fiducial sequence colour over the field, 
including the differential reddening, is at a level of $\pm0.01$ mag, e.g. as between plots (a) and (b) in Fig.~\ref{hst_radius}.
Since we consider the whole cluster field, where this effect is averaged, it gives a fiducial sequence colour offset
at a level of $\pm0.005$ mag.
This offset is even lower in the cases when we use a fiducial sequence presented by the dataset's authors 
after an investigation of the field errors by them.

\section{Theoretical models and isochrones}
\label{iso}

In order to fit the CMDs of NGC\,6205, we use the following theoretical models of stellar evolution and related isochrones:
\begin{itemize}
\item PARSEC together with the COLIBRI~PR16 \citep{marigo2013, rosenfield2016} assuming [Fe/H]$=-1.58$, $Z=0.0004$, 
$Y=0.2485+1.78Z=0.2492$, [$\alpha$/Fe]$=0$, solar $Z=0.0152$, mass loss efficiency $\eta=0.2$, where $\eta$ 
is the free parameter in Reimers' law \citep{reimers};
\item MIST with the average observed photospheric metallicity [Fe/H]$=-1.58$, which corresponds to the initial protostar 
birth cloud bulk metallicity [Fe/H]$=-1.61$, $Z=0.000\,365$, $Y=0.2495$, [$\alpha$/Fe]$=0$ w.r.t. the protosolar 
$Z=0.0142$;
\item DSEP, solar-scaled abundance (hereafter solar-scaled DSEP) with [Fe/H]$=-1.58$, $Z=0.000\,45$, $Y=0.2457$, 
[$\alpha$/Fe]$=0$ and enhanced abundance (hereafter enhanced DSEP) with [Fe/H]$=-1.58$, $Z=0.000\,78$, $Y=0.33$, 
[$\alpha$/Fe]$=+0.40$, both with the solar $Z=0.0189$;
\item BaSTI, solar-scaled abundance (hereafter solar-scaled BaSTI) with [Fe/H]$=-1.614$, $Z=0.0004$, $Y=0.245$, 
[$\alpha$/Fe]$=0$, mass loss efficiency $\eta=0.4$ and enhanced abundance (hereafter enhanced BaSTI) with 
[M/H]$=-1.27$, [Fe/H]$=-1.62$, $Z=0.0009$, $Y=0.3$, [$\alpha$/Fe]$=+0.4$, mass loss efficiency $\eta=0.2$,
both with the solar $Z=0.019$;
\item IAC-BaSTI with [Fe/H]$=-1.58$, $Z=0.0004$, $Y=0.2476$, [$\alpha$/Fe]$=0$, overshooting, diffusion, mass loss 
efficiency $\eta=0.3$
\footnote{See \citet{choi2018} for a discussion of an effect of varying mass loss and other parameters on CMDs.}
and the initial solar $Z=0.0172$ and $Y=0.2695$.
\end{itemize}

The current releases of PARSEC, MIST, and IAC-BaSTI include the solar-scaled models only.

Following \citetalias{ngc5904}, we use only the response curves of $G$, $G_\mathrm{BP}$, and $G_\mathrm{RP}$ from \citet{gaiaevans}, 
but not from \citet{weiler2018}, 
as well as $G$ photometry without a systematic correction proposed by \citet[][their equation (3)]{casagrande2018}.
However, this affects the fitting negligibly.

Following \citet{an2009}, we adjust the SDSS model magnitudes using AB corrections given by \citet{eisenstein2006}:
$u_{AB}=u-0.040$, $i_{AB}=i+0.015$, and $z_{AB}=z+0.030$, with no corrections in the $g$ and $r$ bands.

To derive the best reddening, distance, and age, we calculate the isochrones for a grid of some reasonable 
ages (10--17 Gyr for MIST and IAC-BaSTI or 10--15 Gyr for the remaining models, with a step of 0.5 Gyr), 
distances (5.5--9.5 kpc with a step of 0.1 kpc) and reddenings [between $-0.05$ mag and the value calculated from 
$E(B-V)=0.06$ (the highest reasonable estimate) and the \citetalias{ccm89} extinction law 
with $R_\mathrm{V}=5$ (the highest reasonable value), with a step of 0.005 mag].

Each isochrone is represented by its authors as a discrete number of CMD points: this is seen in our figures with CMDs, 
especially for PARSEC with the most thinned-out isochrone points. 
Therefore, we select the best isochrone of each model as the one with a minimal total offset between the isochrone 
points and the fiducial points in the same magnitude range of the CMD.
To pay more attention to CMD domains with more accurate photometry and well-defined models,
we consider only fiducial points at the TO, SGB, HB, the brighter part of the MS (higher than the magnitude limit applied in 
Sect.~\ref{photo}) and the parts of the RGB and AGB fainter than a CMD point where they meet.

As discussed in \citetalias{ngc5904}, the magnitudes of the HB and SGB better constrain the distance,
the length and slope of the SGB and the magnitude difference between the HB and SGB -- the age,
while the overall colour offset of the isochrone w.r.t. the fiducial sequence -- the reddening.
This is taken into account for the balance of the uncertainties, presented in Appendix~\ref{uncertainties}.

\begin{figure}
\includegraphics{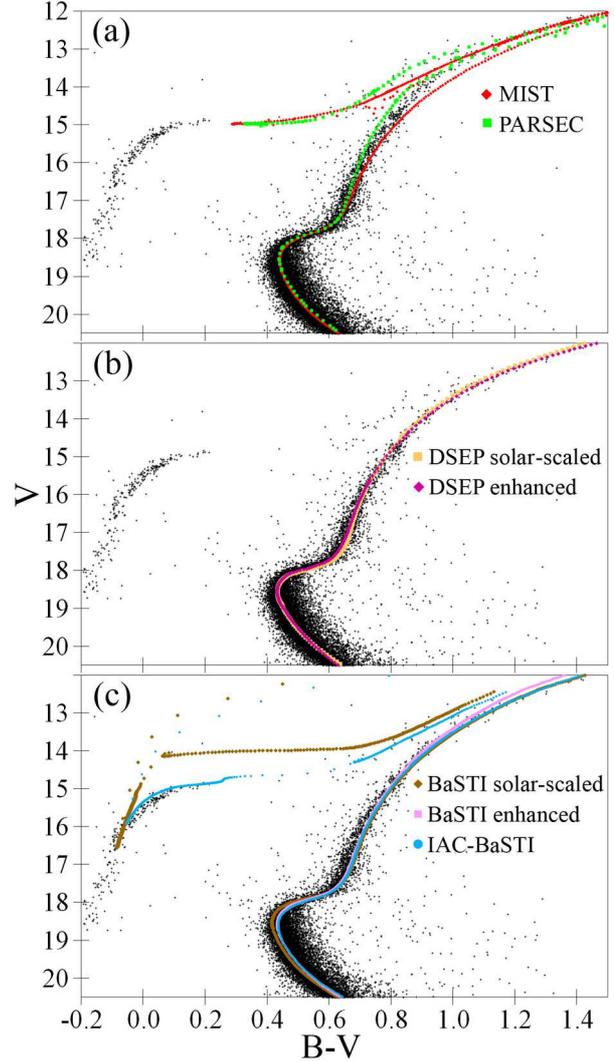}
\caption{$B-V$ versus $V$ CMD of NGC\,6205 based on the data from \citetalias{stetson2019} with the isochrones from
solar-scaled PARSEC (green) and solar-scaled MIST (red) -- upper plot,
solar-scaled DSEP (yellow) and enhanced DSEP (magenta) -- middle plot,
solar-scaled BaSTI (brown), enhanced BaSTI (light purple), and solar-scaled IAC-BaSTI
(blue) -- lower plot, these are calculated with the best-fitting parameters from Table~\ref{fit}.
}
\label{bv}
\end{figure}

\begin{figure}
\includegraphics{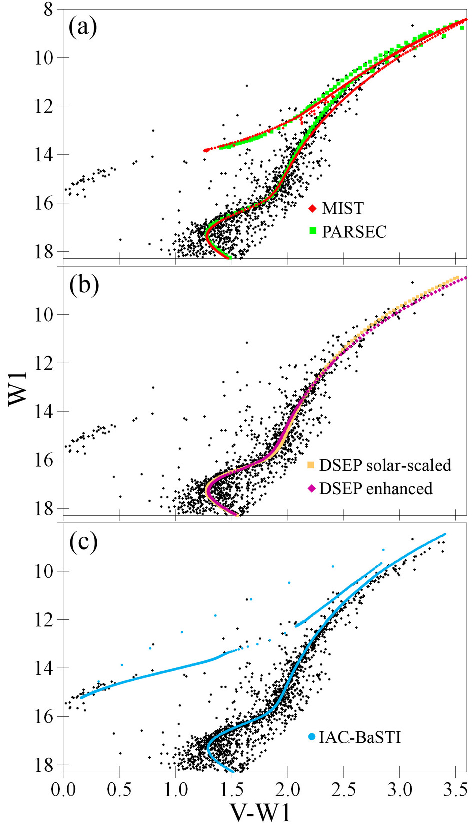}
\caption{The same as Fig.~\ref{bv} but for $V-W1$ versus $W1$ based on the data from \citetalias{stetson2019} and unWISE.
}
\label{vw1}
\end{figure}

\begin{figure}
\includegraphics{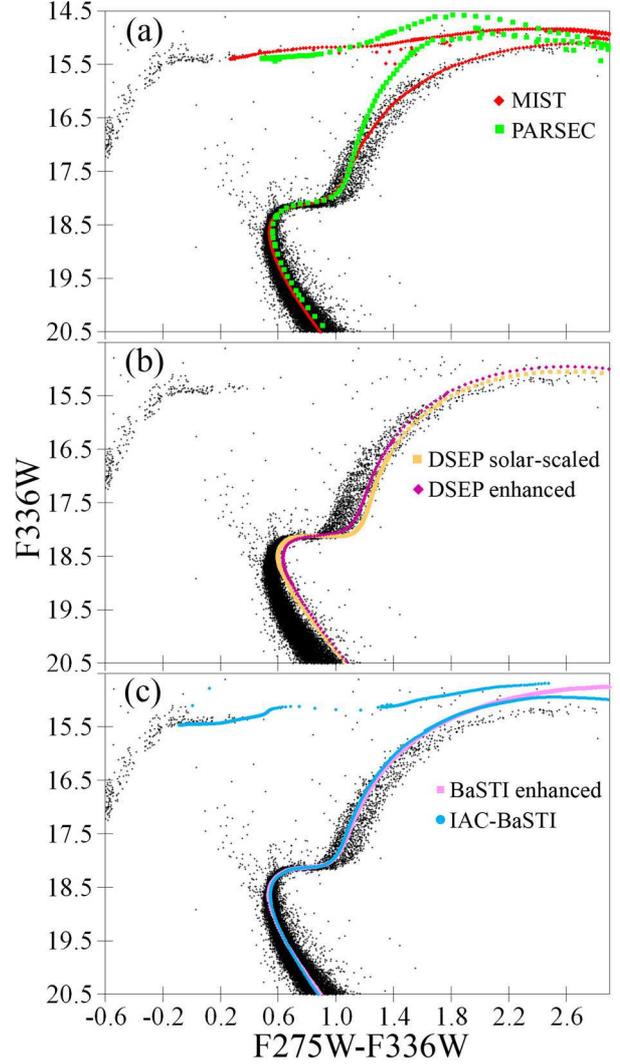}
\caption{The same as Fig.~\ref{bv} but for $F275W-F336W$ versus $F336W$ based on the {\it HST}/WFC3 data.
}
\label{f275_336iso}
\end{figure}

\begin{table*}
\def\baselinestretch{1}\normalsize\normalsize
\caption{The results of the isochrone fitting for various models and some key colours.
To avoid confusion in the cases of the same colour, the source of photometry is indicated in brackets.
The column `Error' contains the predicted reddening uncertainty, which is calculated in Appendix~\ref{uncertainties}. 
In all the CMDs we use the colour as the abscissa and the magnitude in the redder filter as the ordinate: 
for example, $E(B-V)$ with the related age and distance is obtained by use of the $B-V$ versus $V$ CMD.
The complete table is available online.
}
\label{fit}
\[
\begin{tabular}{lcccccccc}
\hline
                                   & PARSEC         & MIST           & Solar-scaled DSEP & Enhanced DSEP  & IAC-BaSTI & Error \\
\hline
$E(B-V)$ \citep{rey2001}           & $0.044\pm0.03$ & $0.055\pm0.03$ & $0.017\pm0.02$    & $0.045\pm0.02$ & $0.013\pm0.02$ & 0.027 \\ 
age, Gyr                           & 12.5           & 14.5           & 14.5              & 12.5           & 15.0           & \\
distance, pc                       & 7600           & 7400           & 7700              & 7200           & 7500           & \\
\hline
$E(B-V)$ \citepalias{stetson2019}  & $0.042\pm0.03$ & $0.060\pm0.02$ & $0.018\pm0.02$    & $0.045\pm0.02$ & $0.020\pm0.02$ & 0.032 \\ 
age, Gyr                           & 13.0           & 15.0           & 15.0              & 12.5           & 14.5           & \\
distance, pc                       & 7400           & 7200           & 7300              & 7200           & 7500           & \\
\hline
$E(V-I)$ \citepalias{stetson2019}  & $0.018\pm0.02$ & $0.046\pm0.02$ & $0.026\pm0.03$    & $0.050\pm0.02$ & $0.017\pm0.02$ & 0.033 \\
age, Gyr                           & 13.5           & 14.5           & 15.0              & 12.5           & 14.0           & \\
distance, pc                       & 7700           & 7600           & 7400              & 7200           & 7800           & \\
\hline
$E(I-W1)$ \citepalias{stetson2019} & $0.065\pm0.02$ & $0.081\pm0.03$ & $0.073\pm0.03$    & $0.076\pm0.03$ & $0.089\pm0.03$ & 0.038 \\
age, Gyr                           & 11.5           & 14.0           & 13.5              & 12.0           & 13.5           & \\
distance, pc                       & 8100           & 8000           & 7600              & 7400           & 7900           & \\
\hline
$E(G_\mathrm{BP}-G_\mathrm{RP})$   & $0.044\pm0.03$ & $0.046\pm0.03$ & $0.053\pm0.02$    & $0.066\pm0.02$ & $0.020\pm0.03$ & 0.034 \\
age, Gyr                           & 12.5           & 14.0           & 13.5              & 12.0           & 14.5           & \\
distance, pc                       & 7800           & 7600           & 7600              & 7200           & 7400           & \\
\hline
$E(G_\mathrm{RP}-W1)$              & $0.089\pm0.02$ & $0.074\pm0.03$ & $0.092\pm0.02$    & $0.120\pm0.02$ & $0.083\pm0.02$ & 0.034 \\
age, Gyr                           & 11.5           & 13.5           & 14.0              & 12.5           & 12.5           & \\
distance, pc                       & 7900           & 7700           & 7500              & 7400           & 8100           & \\
\hline
$E(b-y)$ \citepalias{grundahl1998} & $0.008\pm0.04$ & $0.035\pm0.03$ & $0.035\pm0.03$    & $0.049\pm0.01$  & $0.020\pm0.02$ & 0.039 \\
age, Gyr                           & 13.5           & 14.5           & 14.0              & 11.5            & 14.0           & \\ 
distance, pc                       & 7600           & 7200           & 7000              & 7000            & 7300           & \\
\hline
$E(y-W1)$ \citepalias{savino2018}  & $0.132\pm0.02$ & $0.123\pm0.03$ & $0.102\pm0.03$    & $0.126\pm0.02$  & $0.102\pm0.03$ & 0.039 \\
age, Gyr                           & 12.0           & 14.0           & 13.5              & 12.5            & 13.0           & \\ 
distance, pc                       & 7700           & 7700           & 7500              & 7300            & 7900           & \\
\hline
$E(V-K)$ \citepalias{brasseur2010} & $0.094\pm0.03$ & $0.110\pm0.03$ & $0.083\pm0.03$    & $0.110\pm0.03$  & $0.088\pm0.03$ & 0.034 \\
age, Gyr                           & 13.0           & 15.5           & 14.0              & 13.0            & 14.5           & \\
distance, pc                       & 7200           & 6900           & 7200              & 6900            & 7200           & \\
\hline
\end{tabular}
\]
\end{table*}


\section{Results}
\label{results}

As in \citetalias{ngc5904}, our results for different colours are consistent within their precision.
Therefore, to avoid redundancy in this paper,
we show some key examples of the CMDs with the best isochrone fits in Fig.~\ref{bv}, \ref{vw1} and \ref{f275_336iso} 
(which can be compared with Fig.~\ref{bvv} and \ref{f275_336}) as well as in Fig.~\ref{gbpgrp} -- \ref{vj} in Appendix~\ref{cmds}, 
while the ages, distances, and reddenings derived from some key isochrone fits are presented in Table~\ref{fit}.
Figures and results for all the CMDs can be provided on request.

The predicted uncertainties of the derived distance, age and reddening are described in Appendix~\ref{uncertainties}.
Also, the predicted uncertainties of the derived reddenings are given in the rightmost column `Error' of Table~\ref{fit}.
They can be compared with the maximal offsets of the isochrones w.r.t. the fiducial sequences along the reddening vector
(i.e. nearly along the colour, or the abscissa).
Such an offset is calculated for each isochrone in each CMD within the SGB, TO, brighter part of the MS, 
and fainter parts of the RGB and AGB, and presented in Table~\ref{fit} after each value of reddening.
The offset varies between 0.01 and 0.08 mag, with 0.03 mag as a typical value, and can be accepted as the reddening precision.
As is evident from Table~\ref{fit}, the offsets are comparable with the corresponding predicted uncertainties 
(the `Error' column of Table~\ref{fit}) for the vast majority of isochrones and CMDs. 
This means that we take into account the uncertainties correctly.
For each isochrone in each CMD, the largest value in the pair of the offset and the predicted uncertainty
is adopted as the final accuracy of the derived reddening.
It is used for the analysis in Sect.~\ref{extlaw} and \ref{adjustment}, as well as in Fig.~\ref{law} and \ref{law2}.

The cases with a fiducial--isochrone offset $>0.08$ mag are marked in Table~\ref{fit} as a fail fitting of the 
corresponding CMD.
Also, we have to mark a model fail for a colour, if the best-fitting isochrone provides an improbable noticeable negative reddening 
[corresponding to $E(B-V)<-0.025$ mag].
However, a less noticeable negative or zero reddening is allowed for a CMD due to some errors of the data used or as a 
short-term feature of the observed extinction law.
Fig.~\ref{f275_336iso} presents some examples of such a fail fitting.
In plot (a) the PARSEC isochrone for the distance 7800 pc, age 13 Gyr and reddening $E(F275W-F336W)=0.021$ mag
[about $E(B-V)=0.02$ mag] fails to fit the main part of the RGB, providing too large an isochrone-to-data offset.
A larger reddening would improve this fitting on the RGB, but at the expense of a worse fitting on the fainter RGB and MS.
In plot (b) both the solar-scaled and enhanced DSEP isochrones for the zero reddening fail to fit the MS, TO and RGB,
providing too large an isochrone-to-data offset.
In this case, better fitting can be obtained, but at the expense of unreliable reddening $E(F275W-F336W)<-0.1$ mag.

\begin{table*}
\def\baselinestretch{1}\normalsize\small
\caption[]{The mean and median age (Gyr) and distance (pc) derived for the various models.
}
\label{agedist}
\[
\begin{tabular}{lccccccc}
\hline
\noalign{\smallskip}
 & \multicolumn{7}{c}{Model} \\
\hline
\noalign{\smallskip}
                & PARSEC       & MIST         & DSEP         & DSEP     & BaSTI        & BaSTI    & IAC-BaSTI \\
                &              &              & Solar-scaled & Enhanced & Solar-scaled & Enhanced &           \\
\hline
\noalign{\smallskip}
Mean distance                             & $7630\pm316$ & $7430\pm474$ & $7520\pm368$ & $7220\pm272$  & $7100\pm250$ & $6844\pm260$ & $7462\pm359$ \\
Median distance                           & 7700         & 7400         & 7500         & 7200          & 7100         & 6900         & 7400 \\
Mean distance for $0.45<\lambda<0.9$ nm   & $7545\pm267$ & $7289\pm264$ & $7416\pm279$ & $7147\pm244$  & $7125\pm255$ & $6814\pm212$ & $7314\pm280$ \\
Median distance for $0.45<\lambda<0.9$ nm & 7600         & 7300         & 7500         & 7200          & 7100         & 6900         & 7350 \\
\hline
Mean age                             & $12.6\pm0.9$ & $14.3\pm0.8$ & $13.8\pm1.2$ & $12.1\pm0.8$  & $14.8\pm0.5$ & $14.6\pm0.1$ & $14.1\pm0.7$ \\
Median age                           & 12.5         & 14.0         & 14.0         & 12.5          & 15.0         & 15.0          & 14.0 \\
Mean age for $0.45<\lambda<0.9$ nm   & $12.8\pm0.8$ & $14.4\pm0.7$ & $14.4\pm0.6$ & $12.3\pm0.7$  & $15.0\pm0.0$ & $14.9\pm0.2$  & $14.3\pm0.3$ \\
Median age for $0.45<\lambda<0.9$ nm & 13.0         & 14.5         & 14.5         & 12.5          & 15.0         & 15.0          & 14.5 \\
\hline
\end{tabular}
\]
\end{table*}


\begin{figure*}
\includegraphics{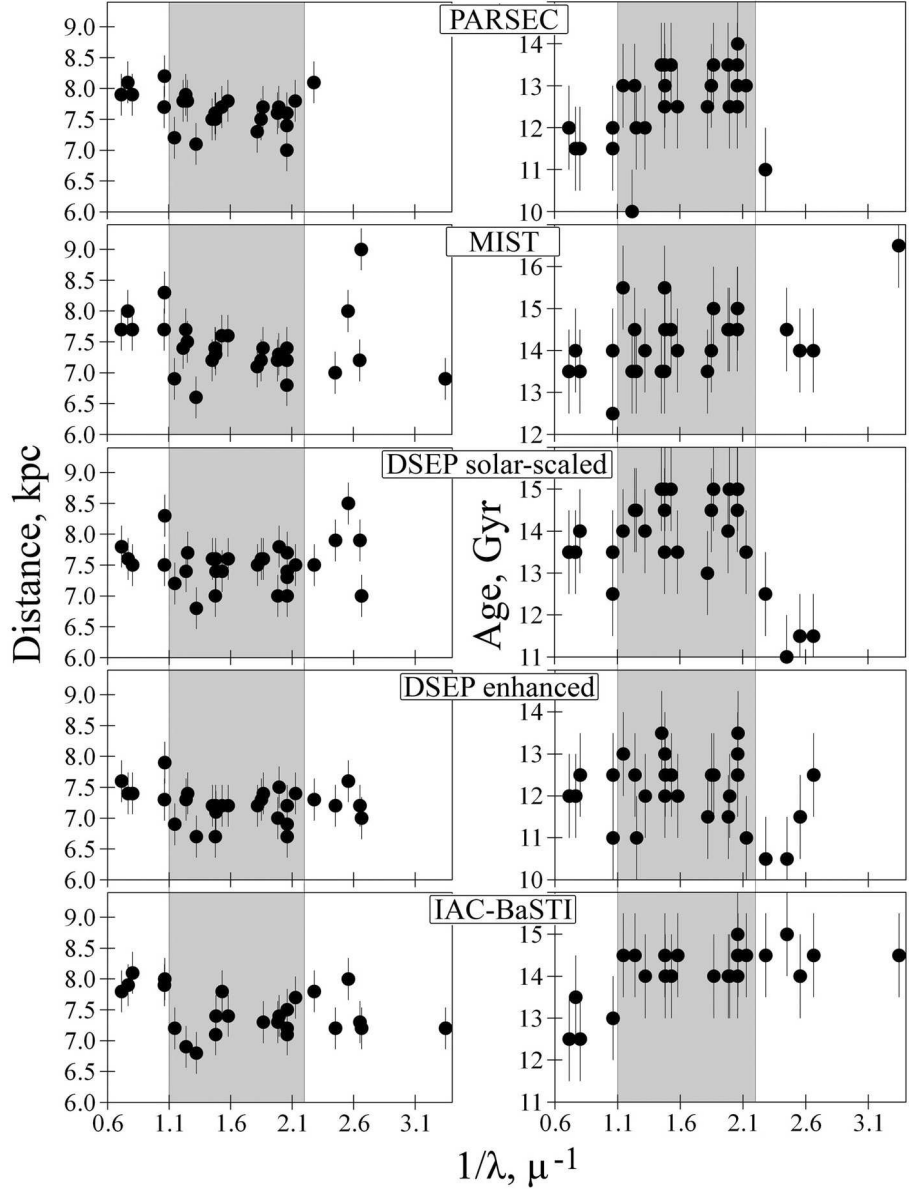}
\caption{The distances and ages, derived by use of the different models, as functions of $1/\lambda$.
The error bars are of 340 pc and 1 Gyr, respectively. 
The grey area is the wavelength range within $0.45<\lambda<0.9$ nm.
}
\label{distage}
\end{figure*}

The age and distance to NGC\,6205 are better determined by use of the models and data for both the HB and SGB 
(PARSEC, MIST and IAC-BaSTI), while DSEP and BaSTI allow us to consider the difference between the solar-scaled and enhanced isochrones.
However, BaSTI gives few isochrones (as seen in the online version of Table~\ref{fit}), with no IR band, 
as well as unreliable ages about 15 Gyr (even older, if BaSTI was not limited to 15 Gyr). 

Table~\ref{agedist} presents the mean and median estimates as well as standard deviations
\footnote{The standard deviation $\pm0.0$ for solar-scaled BaSTI means that all the estimates are 15 Gyr.}
of age and distance from the models.
These values appear different for the UV, optical ($0.45<\lambda<0.9$ nm) and IR ranges
\footnote{Here we calculate $1/\lambda$ for the colour under consideration as an average of $1/\lambda_\mathrm{eff}$ 
for the filters used, where $\lambda_\mathrm{eff}$ is taken from Table~\ref{filters}.}.
This is evident from Fig.~\ref{distage}, where the distances and ages are presented as functions of $1/\lambda$
\footnote{The solar-scaled DSEP isochrones can be calculated only to 15 Gyr, thus, providing only the lower estimate 
for some of its results, as seen from Fig.~\ref{distage}.}.
This issue must be investigated in future.
Assuming that the models and isochrones might be more reliable in the optical range, we use only this range
in order to calculate the most probable estimates of the distance and age.

The different datasets with one model applied provide rather consistent distances and ages,
at least, in the optical range, as is evident from Fig.~\ref{distage} and from the standard deviations in Table~\ref{agedist}.
The predicted uncertainties of the distance and age ($\pm340$ pc and $\pm1$ Gyr, respectively) from 
Appendix~\ref{uncertainties} agree with the standard deviations in Table~\ref{agedist} within the optical range
and slightly underestimate the deviations within the full wavelength range.

However, the distance and age estimates from Table~\ref{agedist} show a considerable systematic discrepancy for the 
different models, which exceeds the predicted uncertainties. This discrepancy exists in the optical range as well.
This follows the high diversity of NGC\,6205 estimates, mentioned in Sect.~\ref{metal}, and the note for age by \citet{saracino2019}:
`Differences in the best-fit age are somewhat expected when different models, which adopt slightly different solar 
abundances, opacities, reaction rates, efficiency of atomic diffusion etc., are compared'.

\begin{figure*}
\includegraphics{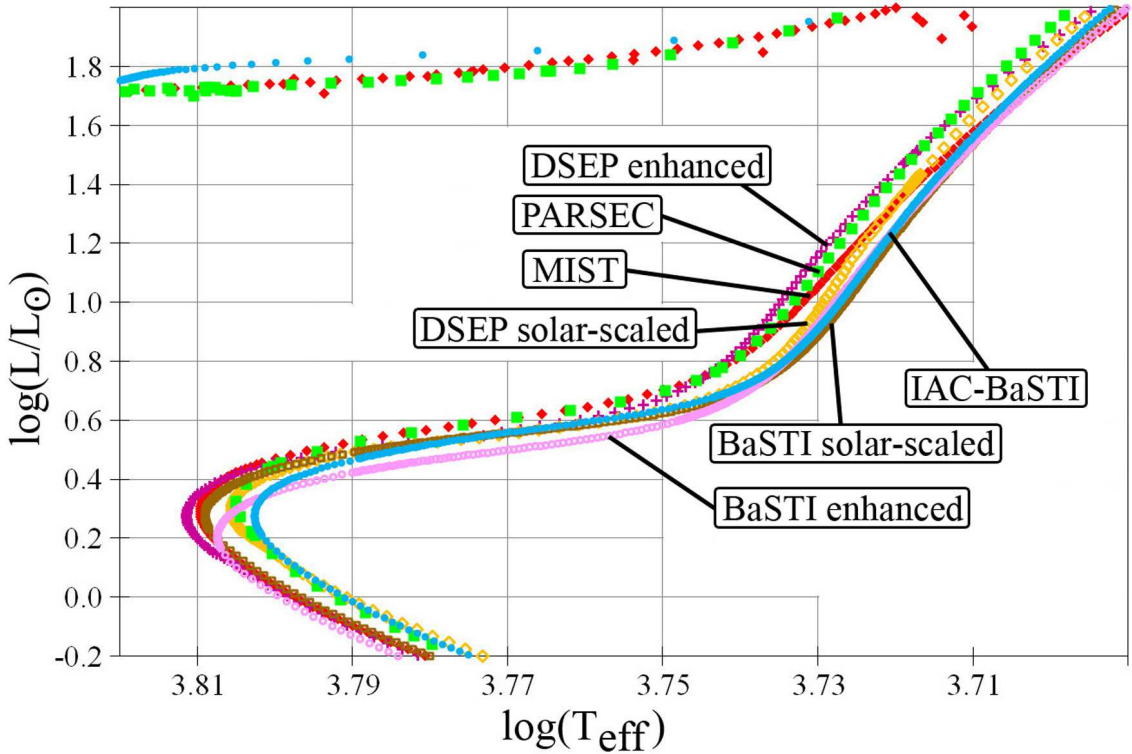}
\caption{$\log(T_\mathrm{eff})$ versus $\log(L/L_{\sun})$ diagram with the isochrones from
PARSEC (green squares), MIST (red diamonds), solar-scaled DSEP (yellow open diamonds), enhanced DSEP (magenta crosses),
solar-scaled BaSTI (brown open squares), enhanced BaSTI (light purple circles), and solar-scaled IAC-BaSTI (blue circles)
for the best-fitting median ages for $0.45<\lambda<0.9$ nm taken from Table~\ref{agedist}.
This figure shows only the stages of stellar evolution used in the isochrone-to-data fitting.
}
\label{teff}
\end{figure*}

Fig.~\ref{teff} presents a diagram of effective temperature $T_\mathrm{eff}$ versus luminosity $L$ 
(w.r.t. the solar one) with the different isochrones.
These isochrones are calculated for the best-fitting median ages taken from Table~\ref{agedist} for the optical range.
A reasonable variation of these ages provides a negligible shift of these isochrones and does not change our conclusions.
Such a diagram is independent of distance, reddening, wavelength range, colour--$T_\mathrm{eff}$ relations and 
bolometric corrections used. Hence, it is a direct comparison of the models.

Fig.~\ref{teff} explains the bulk of the systematic differences between the estimates from different models
in Table~\ref{agedist}.
For example, PARSEC gives the youngest age due to the shortest SGB, while both solar-scaled and enhanced BaSTI give
the oldest ages due to the longest SGB.
IAC-BaSTI gives an age older than PARSEC does due to a larger HB--SGB luminosity difference.
Although this difference for MIST is also small, its long SGB makes the MIST ages rather old.
The derived distances in Table~\ref{agedist} also follow the luminosity of the SGB and TO in Fig.~\ref{teff} 
for all the models:
from the shortest distance and faintest SGB/TO for the enhanced BaSTI 
to the longest distance and brightest SGB/TO for PARSEC.

Extinction and reddening, provided by an isochrone, is primarily defined by an overall (or median) $T_\mathrm{eff}$ of this isochrone.
Fig.~\ref{teff} shows that IAC-BaSTI gives the reddest isochrone and, consequently, needs the lowest reddening 
to fit such an isochrone with data.
BaSTI and solar-scaled DSEP give slightly bluer isochrones, followed by PARSEC, MIST and, finally, by enhanced DSEP with the bluest isochrone.
This explains the lowest extinction/reddening estimates from IAC-BaSTI, higher ones from PARSEC, even higher ones
from MIST and the highest ones from the enhanced DSEP in Sect.~\ref{extlaw} and \ref{adjustment}.

Thus, the diversity of the distance, age, extinction and reddening estimates from the different models 
is mainly due to some differences of the models before the colour--$T_\mathrm{eff}$ relations and bolometric corrections 
are applied.

In Fig.~\ref{bv} -- \ref{f275_336iso} and other CMDs, BaSTI and DSEP show quite little (a few hundredths of a magnitude) difference 
between the best-fitting solar-scaled and enhanced isochrones.
However, Table~\ref{fit} and \ref{agedist} show that the derived age, reddening and distance, for the best-fitting solar-scaled and 
enhanced isochrones in the same CMD, differ considerably.

Fig.~\ref{teff} shows that the enhancement makes both the DSEP and BaSTI isochrones fainter, 
which is followed by a shorter distance for the enhanced models.
However, the enhancement decreases the age both for DSEP and BaSTI, despite a negligible change of the length of the SGB,
as seen from Fig.~\ref{teff}.
Therefore, this age decrease may be due to the colour--$T_\mathrm{eff}$ relations and/or bolometric corrections.
Also, Fig.~\ref{teff} shows that the enhanced DSEP isochrone is much bluer than the solar-scaled one, 
while there is no such a difference for BaSTI.
Thus, the effect of enhancement on the derived parameters is not clear.

We decide to use only PARSEC, MIST, enhanced DSEP and IAC-BaSTI median estimates for $0.45<\lambda<0.9$ nm
in the calculation of the most probable age and distance.

The most probable distance for NGC\,6205 is $7.4\pm0.2$ kpc, as a middle value for the interval 7.2--7.6 kpc 
of the PARSEC, MIST, enhanced DSEP and IAC-BaSTI estimates. The uncertainty includes both random and systematic uncertainties.
The correspondent true distance modulus is $(m-M)_0=14.35\pm0.06$ mag, while the parallax is $0.135\pm0.004$ mas.
This agrees within the uncertainties with recent estimate of the parallax from {\it Gaia} DR2 
$0.0801+0.029=0.1091\pm0.040$ mas, where $0.029$ and $0.040$ mas are the parallax global zero-point correction 
derived from the analysis of the high-precision quasar sample and the median uncertainty in parallax, 
respectively \citep{lindegren2018}. This {\it Gaia} DR2 parallax corresponds to the distance $R=9.2^{+5.3}_{-2.5}$ kpc.

The age of NGC\,6205 remains quite uncertain until the development of some new enhanced models.
We assume with caution that the most probable age for NGC\,6205 is 13.5 Gyr, as the middle value for the interval 
12.5--14.5 Gyr of the PARSEC, MIST, enhanced DSEP and IAC-BaSTI estimates, 
with an uncertainty of 1 Gyr covering the whole interval of these estimates.

In \citetalias{ngc5904} we have shown with similar data and using a similar approach that any reasonable variation 
of metallicity, abundance, age and distance cannot change an isochrone colour. 
Hence, such a colour w.r.t. a fiducial sequence provides us with an estimate of the reddening.
Each pair of a model and a dataset provides its own set of reddenings for different wavelengths
and, consequently, its own extinction law. 
Below, we compare them with each other and with the most popular extinction laws.

\subsection{Preliminary extinction law}
\label{extlaw}

Table~\ref{fit} shows that \citetalias{stetson2019} and \citet{rey2001}, in combination with PARSEC, MIST, enhanced DSEP and IAC-BaSTI,
give an agreed direct estimate of the average $E(B-V)=0.04$ mag with both the random and systematic uncertainty 0.01 mag, 
assuming that systematic errors of the different models/isochrones compensate each other.

PARSEC, MIST, DSEP and IAC-BaSTI, in contrast to BaSTI, provide us with IR colours.
As discussed in \citetalias{ngc5904}, the IR extinction and its variation due to some reasonable variations of the 
extinction law are very low.
For example, $E(B-V)=0.04$ mag from our results with the \citetalias{ccm89} extinction law with its reasonable variations
within $2.6<R_\mathrm{V}<3.6$
\footnote{\citetalias{ccm89} do not provide $A_\mathrm{W1}$, but it is extrapolated by us from the \citetalias{ccm89} IR extinction law.}
corresponds to $0.011<A_\mathrm{K}<0.018$ and $0.006<A_\mathrm{W1}<0.009$ mag.
Such estimates would be two times less with $E(B-V)=0.02$ from \citet{harris}.
Hence, we can initially adopt $A_\mathrm{K}=0.012\pm0.006$ and $A_\mathrm{W1}=0.006\pm0.003$ mag, where both the uncertainties of the 
extinction and extinction law are included.
Thus, the uncertainty of the IR extinction is several thousandths of a magnitude.
This is a minor contribution into the balance of the uncertainties discussed in Appendix~\ref{uncertainties}.

We calculate the extinction in the remaining 32 bands, by use of these IR extinctions calculated with $R_\mathrm{V}=3.1$, 
in combination with the reddenings derived from the isochrone fitting.
In such an approach, we have to adopt an extinction law only for the first iteration, i.e. to calculate $A_\mathrm{K}$ and $A_\mathrm{W1}$.
Then we derive $R_\mathrm{V}$ from the extinctions in the remaining bands.
In the next iteration we recalculate $A_\mathrm{K}$ and $A_\mathrm{W1}$ by use of the \citetalias{ccm89} extinction 
law with the derived $R_\mathrm{V}$. 
Since $R_\mathrm{V}\approx3.1$ for NGC\,6205, as shown later, we need only a few iterations to converge.
The derived $A_\mathrm{K}\approx0.014$ and $A_\mathrm{W1}\approx0.007$ (slightly different for the different models)
define the IR extinction zero-points for the datasets with IR observations, as mentioned in Sect.~\ref{photo}.

Our reddening results do not contradict previous results by other authors on a short wavelength baseline. For example,
the average $E(b-y)=0.028\pm0.018$ mag from PARSEC, MIST, enhanced DSEP and IAC-BaSTI in Table~\ref{fit} does not 
contradict the estimate $E(b-y)=0.015\pm0.010$ by \citetalias{grundahl1998} for the same data but using a different method,
however, our estimate is almost twice as high.
However, a longer wavelength baseline provides us with a more precise reddening/extinction estimate and also 
smoothes possible short-term variations of the extinction law.

The reddening $E(B-V)$ represents the extinction law only on a short wavelength baseline, while the extinction 
$A_\mathrm{V}$ is well defined on a much longer baseline as 
\begin{equation}
\label{avaw1}
A_\mathrm{V}\approx A_\mathrm{V}-A_\mathrm{W1}=E(V-W1).
\end{equation}
Therefore, given the estimates for $A_\mathrm{K}$ and $A_\mathrm{W1}$, we calculate $A_\mathrm{V}$ from equation~(\ref{avaw1}), 
or from similar equations with $K$ instead of $W1$ or $y$ instead of $V$, for each model by averaging the 
isochrone fitting results for the datasets of (i) \citetalias{stetson2019}, (ii) \citetalias{brasseur2010}, 
(iii) \citetalias{savino2018} and \citetalias{grundahl1998}.
We obtain preliminary $A_\mathrm{V}=0.112$, 0.131, 0.132, and 0.106 mag for PARSEC, MIST, enhanced DSEP, and IAC-BaSTI, 
respectively, which will be adjusted in Sect.~\ref{adjustment}.
The uncertainty of these values can be estimated as $\pm0.02$ mag. The average of these estimates is 
$A_\mathrm{V}=0.120\pm0.014$ mag.

Given $A_\mathrm{V}=0.120\pm0.014$ and $E(B-V)=0.04\pm0.01$ mag, we obtain $R_\mathrm{V}=3.0^{+1.5}_{-0.9}$.
Since this value is quite uncertain and is not significantly different from the value $R_\mathrm{V}=3.1$, 
we decide to keep the latter in our further calculations.

\begin{figure*}
\includegraphics{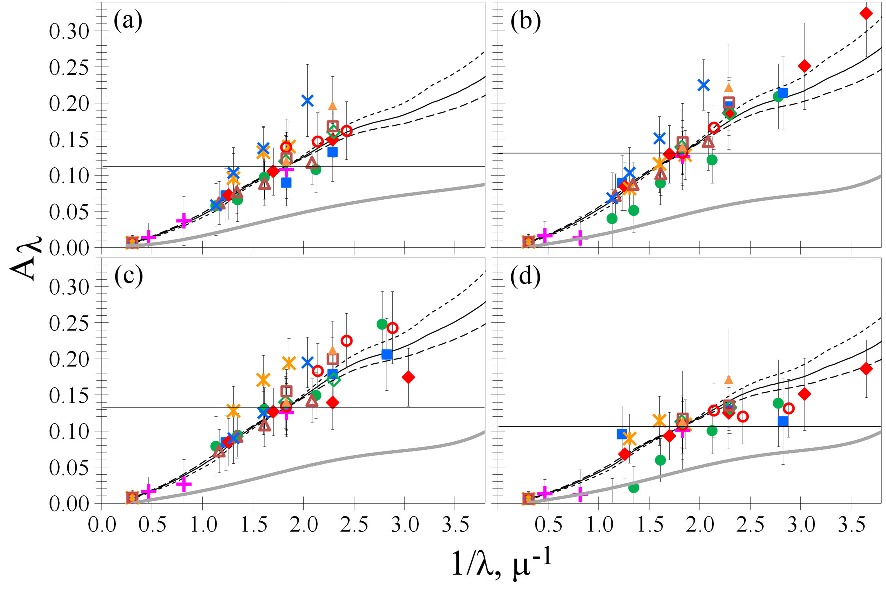}
\caption{The empirical extinction laws from the isochrone fitting by (a) PARSEC, (b) MIST, (c) enhanced DSEP, 
(d) IAC-BaSTI. The datasets with an IR extinction zero-point:
$UBVI$ by \citetalias{stetson2019} -- blue squares, $BV$ by \citet{rey2001} -- open brown squares,
{\it Gaia} -- yellow snowflakes, SDSS -- green circles, Str\"omgren by \citetalias{savino2018} and 
\citetalias{grundahl1998} -- open red circles, $VJK$ by \citetalias{brasseur2010} -- purple upright crosses, and
the datasets with a zero-point defined by the \citetalias{ccm89} extinction law with $R_\mathrm{V}=3.1$:
{\it HST} ACS/WFC3 -- red diamonds, {\it HST} WFPC2 -- open green diamonds,
$BV$ by \citetalias{paltrinieri1998} -- yellow triangles, Pan-STARRS -- open brown triangles,
CFHT/MegaCam by \citet{clem2008} -- blue inclined crosses.
The black dotted, solid and dashed curves show the \citetalias{ccm89} extinction law with $R_\mathrm{V}=2.6$, 3.1, 
and 3.6, respectively, and with the derived $A_\mathrm{V}$, which is shown by the horizontal line.
The thick grey curve shows the extinction law of \citet{schlafly2016} with $A_\mathrm{V}=0.044$ mag.
}
\label{law}
\end{figure*}

Fig.~\ref{law} shows the empirical extinction laws derived from the isochrone fitting for different models and datasets.
The different datasets are shown by different colours.
The uncertainties of the extinctions are calculated from the uncertainties of the reddenings and shown in Fig.~\ref{law} 
by the vertical bars.
Fig.~\ref{law} shows a moderate agreement between the empirical extinction laws from the different models and datasets: 
a smooth extinction law curve can be drawn through all the results within their uncertainty bars. 
However, some noticeable offsets are seen for some datasets, such as the SDSS and {\it Gaia} ones.

The isochrone fitting provides us with the reddenings, which are seen in Fig.~\ref{law} as the relative vertical positions of 
the points. 
Hence, our result is an overall tilt of the bulk of the points in Fig.~\ref{law}.
Therefore, given the fitting results, we cannot shift vertically the points w.r.t. each other.
Also, we cannot shift down the bulk of the points, as a whole, due to a non-negative $A_\mathrm{W1}$.

The thick grey curve in Fig.~\ref{law} shows the extinction law of \citet{schlafly2016} with $E(B-V)=0.015\pm0.001$ and 
$A_\mathrm{V}=0.044\pm0.003$ taken from \citet{schlaflyfinkbeiner2011} [recalibrated 2D reddening map of \citet[][hereafter SFD]{sfd} 
with $E(B-V)=0.017$ and $A_\mathrm{V}=0.051$ mag]
\footnote{Also \citet{schlaflyfinkbeiner2011} find a negligible differential reddening in the field of NGC\,6205 within 
$0.012<E(B-V)<0.017$ mag.}.
We do not draw other popular extinction laws, such as, those of \citetalias{ccm89}, \citet{fitzpatrick1999} and \citet{wang2019} 
with the same low $A_\mathrm{V}$, since they would be indistinguishable from the law of \citet{schlafly2016} 
in the scale of Fig.~\ref{law}, except for their slight deviation in the UV.
However, using three black curves we depict the \citetalias{ccm89} extinction law with the derived high $A_\mathrm{V}$ and different $R_\mathrm{V}$.

It is seen from Fig.~\ref{law} that the grey curve of \citet{schlafly2016}'s extinction law is very far from our results 
and from all three curves of the \citetalias{ccm89} extinction law.
The proximity of the three \citetalias{ccm89} extinction laws with each other means that $R_\mathrm{V}$ does not matter and can be 
determined from our data with a low accuracy only.
In contrast, the deviation of the \citetalias{ccm89} extinction laws with the high $A_\mathrm{V}$ from \citet{schlafly2016}'s
extinction law with the low $A_\mathrm{V}$ means that $A_\mathrm{V}$ does matter and can be determined from these data fairly accurately.

We emphasize the key inconsistency between our average estimate $A_\mathrm{V}=0.120\pm0.014$ mag 
(or even our lowest estimate $A_\mathrm{V}=0.106\pm0.020$ mag from IAC-BaSTI)
and that of \citet{schlaflyfinkbeiner2011}, $A_\mathrm{V}=0.044\pm0.003$ mag, both determined on some long wavelength baselines.
Such long baselines guarantee that this inconsistency cannot be resolved by a reasonable variation of the extinction law.
Thus, all the datasets and models presented in Fig.~\ref{law} show that $A_\mathrm{V}$ from \citet{schlaflyfinkbeiner2011} and,
consequently, from \citetalias{sfd} as the original source, may be underestimated.
Our results seem to be more reliable, since they are based on the multiband photometry on a long wavelength baseline,
while the \citetalias{sfd} reddening estimate is based on some measurements of IR emission and emission-to-reddening calibration.
Hence, the possible underestimation of the reddening by \citetalias{sfd} may be due to a wrong
emission-to-reddening calibration of low reddenings in \citetalias{sfd}, as discussed, for example, by \citet{gm2018}.

\section{Adjustment}
\label{adjustment}

\begin{figure*}
\includegraphics{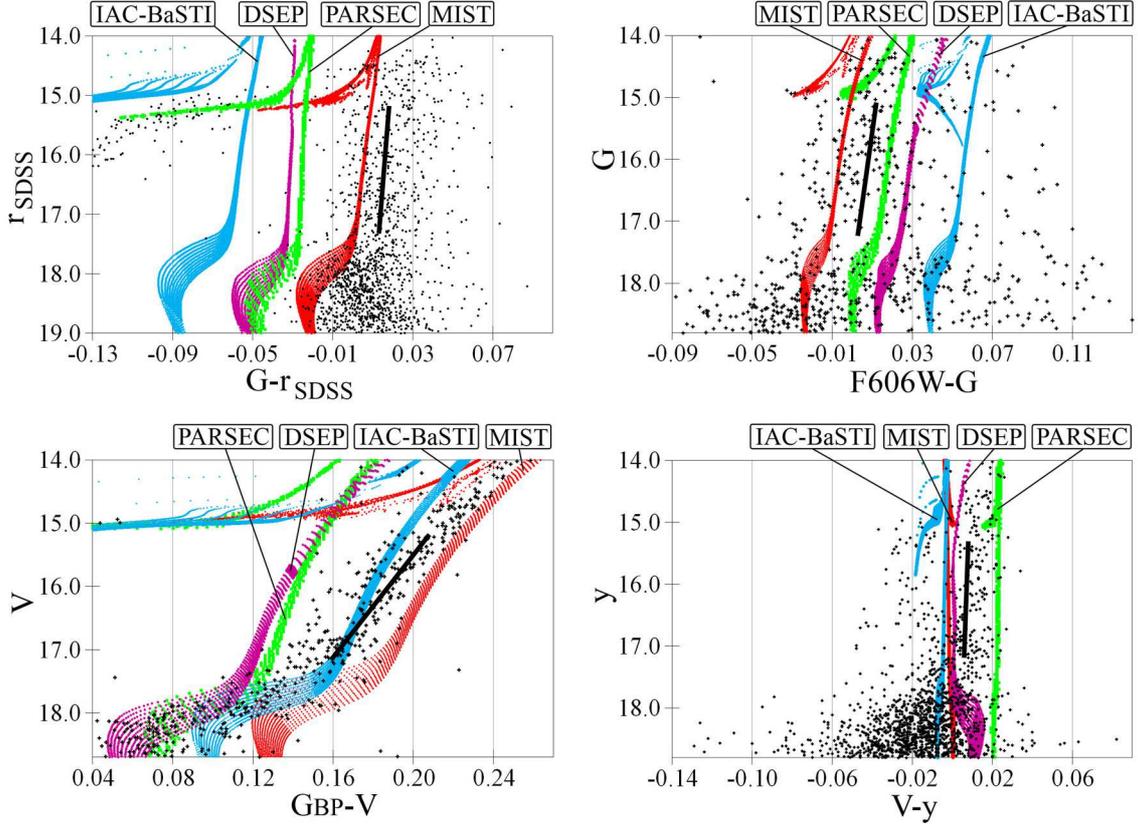}
\caption{Some CMDs of NGC\,6205 for pairs of filters from different datasets but of similar effective wavelengths.
The individual stars are shown by the black dots. 
Part of the fiducial sequence for the RGB between the HB and the SGB is shown by the thick black line.
The sets of isochrones for the ages from 10--15 Gyr, distance 7.4 kpc, reddening $E(B-V)=0.04$~mag and 
\citetalias{ccm89} extinction law with $R_V=3.1$ are shown by the dotted curves and coloured as follows:
PARSEC -- green; MIST -- red; enhanced DSEP -- magenta; and IAC-BaSTI -- blue.
}
\label{offset}
\end{figure*}

Some inherent offsets of the model colours, being different for different datasets, may explain the vertical offsets 
(i.e. the different extinction laws) of the datasets, which are seen in Fig.~\ref{law}.

Similar to \citetalias{ngc5904}, to verify this assumption, we analyse some CMDs obtained with pairs of filters from different 
datasets but of similar effective wavelengths.
Such filters must have a difference of $<0.2$ in their extinction coefficients.
In such a case, any reasonable variation of the reddening [e.g. within $0.01<E(B-V)<0.06$ mag]
cannot shift any isochrone in such a CMD along the colour by more than $0.2\times(0.06-0.01)=0.01$ mag.
This is smaller than the colour offsets under consideration.
Any reasonable variation of [Fe/H], abundance, age, and distance also cannot change these isochrone colours by more than 0.01 mag.
This allows us to consider any colour offset between an isochrone and the fiducial sequence in such a CMD as a systematic error (offset) 
of this isochrone.

Four such CMDs with isochrones from PARSEC, MIST, enhanced DSEP,
\footnote{The solar-scaled DSEP isochrones are not shown, since in all these CMDs and for all the ages they 
coincide within 0.01 mag with the enhanced DSEP isochrones.},
and IAC-BaSTI for the age $10-15$ Gyr, distance 7.4 kpc, $E(B-V)=0.04$ and \citetalias{ccm89} extinction law with $R_V=3.1$ 
are presented in Fig.~\ref{offset}.
Also, we use the $F438W-B$ versus $B$, $G_\mathrm{BP}-y$ versus $y$, and $i_\mathrm{SDSS}-G_\mathrm{BP}$ versus $G_\mathrm{BP}$ CMDs,
which are not shown in Fig.~\ref{offset}.
Thus, we only use the SDSS, {\it Gaia} DR2, \citetalias{stetson2019}, {\it HST}, and Str\"omgren datasets.

In such a CMD, any fiducial sequence at the brightest and faintest magnitudes would be biased due to the sample incompleteness 
and larger photometric errors.
In addition, the HB, AGB and RGB may be confused with each other, since the colours are close to zero.
Therefore, we draw in Fig.~\ref{offset} and use parts of the fiducial sequence for the fainter RGB, i.e. within a magnitude range 
between the HB and the SGB.
It is seen that the scatter of the star colours in this magnitude range is a few hundredths of a magnitude,
i.e. in a good agreement with the stated precision of the photometry.
Moreover, it is evident from Fig.~\ref{offset}
that part of any fiducial sequence in this magnitude range must be very close to a straight line.
Such parts of the fiducial sequences are shown in Fig.~\ref{offset} by the thick black lines.

Also, it is evident from Fig.~\ref{offset} that the colour offsets between these fiducial sequence lines and the parts of the isochrones
within the same magnitude range can be derived with an accuracy better than 0.01 mag.

Fig.~\ref{offset} shows that these colour offsets are different for different isochrones in the same CMD, as well as for different colours.
This may be explained, among other reasons, by discordant colour--$T_\mathrm{eff}$ relations and bolometric corrections 
used with each model for the different datasets, as discussed by \citet{paxton2011, choi2016, newbasti}.
Hence, these offsets can be used to simultaneously adjust the datasets: we search for some consistent colour corrections to shift 
each isochrone to the correspondent fiducial sequence line.
This brings together the extinction estimates from a model for the different datasets, i.e. the datasets within each plot of Fig.~\ref{law}.
We adjust only the datasets, but not the models/isochrones w.r.t. each other, since we have no indication, which model/isochrone is better.
Consequently, an
additional condition of this adjustment is a minimal (nearly zero) absolute value of a sum of the corrections to each model/isochrone.
However, since only five datasets are used in this adjustment, we do not fix the average extinctions of the models,
in contrast to \citetalias{ngc5904}.
However, since these five datasets are the major ones, the average extinctions change slightly.
Also, since we adjust only the colours, this does not change the derived ages and distances.

\begin{table}
\def\baselinestretch{1}\normalsize\small
\caption[]{The offset corrections (mag) applied to the datasets and models.
}
\label{tableoffset}
\[
\begin{tabular}{lrrrr}
\hline
\noalign{\smallskip}
 & \multicolumn{4}{c}{Model} \\
\noalign{\smallskip}
 Dataset & PARSEC & MIST & Enhanced DSEP & IAC-BaSTI \\
\hline
\noalign{\smallskip}
SDSS                            & $-0.02$  & $-0.02$  & $-0.01$ & $-0.04$    \\
{\it Gaia} DR2                  &  $0.03$  & $-0.01$  & $0.03$  & $0.03$    \\
\citetalias{stetson2019}        & $-0.01$  & $0.02$   & 0.00    & $0.02$    \\
{\it HST}                       &  $0.02$  & $0.01$   & $-0.01$ & $-0.02$     \\
Str\"omgren                     & $-0.01$  & $0.00$   & $-0.01$ & $0.00$     \\
\hline
\end{tabular}
\]
\end{table}


\begin{figure*}
\includegraphics{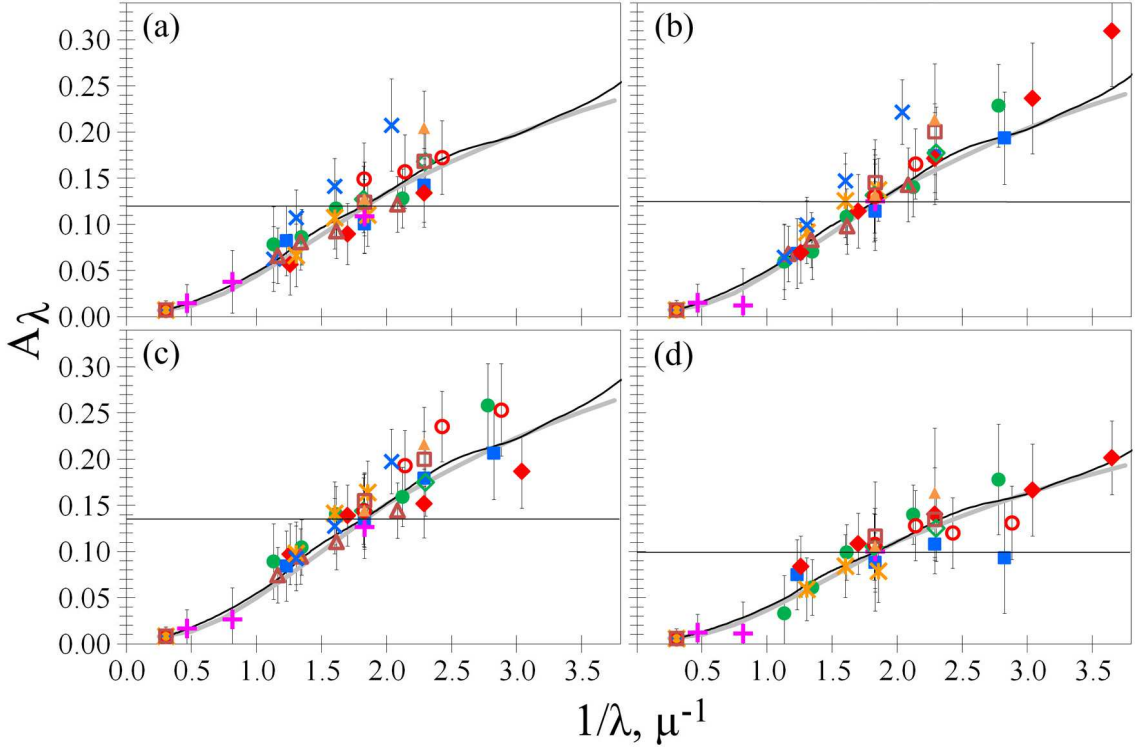}
\caption{The empirical extinction laws from the isochrone fitting by (a) PARSEC, (b) MIST, (c) enhanced DSEP, 
(d) IAC-BaSTI after the adjustment for the corrections from Table~\ref{tableoffset}.
The datasets are:
$UBVI$ by \citetalias{stetson2019} -- blue squares,
$BV$ by \citet{rey2001} -- open brown squares,
{\it Gaia} -- yellow snowflakes,
SDSS -- green circles,
Str\"omgren by \citetalias{savino2018} and \citetalias{grundahl1998} -- open red circles,
$VJK$ by \citetalias{brasseur2010} -- purple upright crosses,
{\it HST} ACS/WFC3 -- red diamonds,
{\it HST} WFPC2 -- open green diamonds,
$BV$ by \citetalias{paltrinieri1998} -- yellow triangles,
Pan-STARRS -- open brown triangles,
CFHT/MegaCam by \citet{clem2008} -- blue inclined crosses.
The datasets without an IR extinction zero-point have a zero-point defined by the \citetalias{ccm89} extinction law with $R_\mathrm{V}=3.1$.
The thick grey and black solid curves show the extinction laws of \citet{schlafly2016} and \citetalias{ccm89} 
with $R_\mathrm{V}=3.1$, respectively, with the derived $A_\mathrm{V}$, which is shown by the horizontal line.
}
\label{law2}
\end{figure*}

\begin{table*}
\def\baselinestretch{1}\normalsize\normalsize
\caption[]{The estimates of $A_\mathrm{V}$ from the various datasets and models.
}
\label{avest}
\[
\begin{tabular}{lrrrrr}
\hline
\noalign{\smallskip}
 & \multicolumn{4}{c}{Model} \\
\noalign{\smallskip}
 Dataset & PARSEC & MIST & Enhanced DSEP & IAC-BaSTI & Average \\
\hline
\noalign{\smallskip}
\citetalias{stetson2019}, \citetalias{brasseur2010}, \citetalias{savino2018}                   & $0.120\pm0.026$ & $0.124\pm0.008$ & $0.135\pm0.009$ & $0.099\pm0.010$ & $0.120\pm0.021$ \\
\citetalias{stetson2019}, \citetalias{brasseur2010}, \citetalias{savino2018}, {\it Gaia}, SDSS & $0.118\pm0.019$ & $0.126\pm0.008$ & $0.142\pm0.012$ & $0.099\pm0.013$ & $0.121\pm0.022$ \\
\hline
\end{tabular}
\]
\end{table*}


The resulting offset corrections applied to the datasets and models are shown in Table~\ref{tableoffset}. 
They demonstrate a consistency with each other and with the colour offsets in Fig.~\ref{offset}.
This adjustment seems to reduce the systematic offsets of the used datasets to a level of $<0.01$ mag.

Fig.~\ref{law2} shows the empirical extinction laws after the adjustment.
It is seen that the adjustment results in a better agreement of the datasets for each model, while there is a slightly 
worse agreement between the different models.
This residual disagreement of the models seems to reflect their inherent systematic differences, 
when the systematic differences between the datasets are removed.

The disagreement of the models in the $E(B-V)$ estimates in Table~\ref{fit} by use of the \citetalias{stetson2019} dataset is $\pm0.02$ mag.
Similar values are obtained from the scatter of the different isochrones along $T_\mathrm{eff}$ in Fig.~\ref{teff}:
it varies from $\pm70$ K at the TO to $\pm50$ K at the RGB and MS, which corresponds to uncertainties of the $B-V$ colour of 
$\pm0.02$ and $\pm0.015$ mag, respectively \citep{casagrande2014}.
Thus, the residual disagreement of the extinction laws given by the different models in Fig.~\ref{law2} can be explained 
by some inherent differences between the models.

A similar residual disagreement of the models is seen in a disagreement of the $A_\mathrm{V}$ estimates, which change slightly after 
the adjustment.
These estimates are presented in Table~\ref{avest} as average values from the datasets of \citetalias{stetson2019}, 
\citetalias{brasseur2010}, and \citetalias{savino2018} (adopting $A_\mathrm{y}=A_\mathrm{V}$ with enough accuracy),
as well as average values with the addition of the datasets of {\it Gaia} and SDSS.
For them, we adopt the relations 
$A_\mathrm{V}=0.52(A_\mathrm{G_\mathrm{BP}}+A_\mathrm{G})$ and 
$A_\mathrm{V}=0.48(A_\mathrm{g}+A_\mathrm{r})$,
which are close enough to any reliable extinction law.
It is seen from Table~\ref{avest} that both the estimates, with or without the additional datasets, give us the robust final result
$A_\mathrm{V}=0.12\pm0.02$. This is the main result of our study.
Since it is based on equation (\ref{avaw1}) or its analogues, this estimate of $A_\mathrm{V}$ is independent of 
the \citetalias{ccm89} or any other extinction law.

$A_\mathrm{V}=0.12\pm0.02$ and $(m-M)_0=14.35\pm0.06$ mag corresponds to an apparent $V$-band distance modulus of 
$(m-M)_\mathrm{V}=14.47\pm0.07$ mag.
This agrees with 
$(m-M)_\mathrm{V}=14.42$ mag from \citet{harris}, $(m-M)_\mathrm{F606W}=14.45\pm0.05$ mag from \citet{vandenberg2013}, and
$(m-M)_\mathrm{V}=14.442^{+0.006}_{-0.005}$ from \citet{wagner-kaiser2017}.

Our final consistent estimates are: $A_\mathrm{V}=0.12\pm0.02$, $E(B-V)=0.04\pm0.01$, and $R_\mathrm{V}=3.1^{+1.6}_{-1.1}$.
Note that such an uncertain $R_\mathrm{V}$ is expected, given the uncertainties of the extinctions in Fig.~\ref{law2}.

The extinction laws of \citet{schlafly2016} and \citetalias{ccm89} with the derived $A_\mathrm{V}$ and $R_\mathrm{V}=3.1$ are shown 
in Fig.~\ref{law2}.
The curves for the laws of \citet{fitzpatrick1999} and \citet{wang2019} would be indistinguishable from that 
of \citet{schlafly2016} on the scale of Fig.~\ref{law2}, except for their slight deviation in the UV.
All the laws agree with our results.

Similar to NGC\,5904 in \citetalias{ngc5904}, we have found a rather high optical extinction for NGC\,6205. 
It is about twice as high than the `canonical' $A_\mathrm{V}=0.06$ mag [as a product of $E(B-V)=0.02$ by $R_\mathrm{V}=3.1$] 
from \citet{harris}, in perfect agreement with the recent estimate of \citet{wagner-kaiser2017} mentioned in Sect.~\ref{intro}.
As is evident from Fig.~\ref{law2} and Table~\ref{fit}, this high extinction is due to rather high reddenings between the IR 
and optical bands.

\section{Conclusions}
\label{conclusions}

This study generally follows \citetalias{ngc5904} in its approach to estimate distance, age and extinction law of a Galactic GC from
a fitting of the isochrones to a multiband photometry. Now we consider NGC6205 (M13) instead of NGC5904 (M5) as in \citetalias{ngc5904}.
We used the photometry of NGC\,6205 in 34 filters from the {\it HST}, unWISE, {\it Gaia} DR2, SDSS, Pan-STARRS,
\citetalias{stetson2019}'s $UBVI$ and other datasets.
These filters cover a wavelength range from about 235 to 4070\,nm, i.e. from the UV to mid-IR.
Some of the photometric datasets were cross-identified with each other and with the unWISE catalogue.
This allowed us to use the photometry in IR bands with nearly zero extinction as an IR extinction zero-point.

We accept the metallicity [Fe/H]$=-1.58$ based on the spectroscopy taken from the literature.

To fit the data, we used the following seven theoretical models of the stellar evolution:
PARSEC, MIST, DSEP (the solar-scaled and He--$\alpha$--enhanced), 
BaSTI (the solar-scaled and He--$\alpha$--enhanced), and IAC-BaSTI.

From the fitting of the isochrones to the fiducial sequences we found the distance, age and reddening estimates for 
each pair of a colour and an isochrone in a CMD.
By use of the reddening estimates we obtained empirical extinction laws, one law for each combination of a model and a dataset.

The derived empirical extinction laws, averaged over the models, agree with the laws of \citet{fitzpatrick1999}, \citet{schlafly2016}, 
\citet{wang2019}, and \citetalias{ccm89} with the best-fitting $R_\mathrm{V}=3.1^{+1.6}_{-1.1}$.

The fitting revealed that for NGC\,6205, with a dominant population enhanced both in helium and in $\alpha$ elements, 
the difference between solar-scaled and enhanced isochrones is quite small, and both kinds of isochrones can be used.

The CMDs, obtained with pairs of filters from different datasets but of similar effective wavelengths, show some colour offsets up 
to 0.04 mag between the fiducial sequences and isochrones.
We attribute these offsets to systematic differences of the datasets.
However, some intrinsic systematic differences of the models/isochrones remain in our results:
the derived distances and ages are systematically different for the ultraviolet, optical and infrared photometry used, 
as well as for the different models/isochrones.
The latter is explained by the differences at a level of $\Delta T_\mathrm{eff}=\pm50$ K [i.e. $\Delta(B-V)=\pm0.015$ mag] 
between the isochrones in the $T_\mathrm{eff}$ -- luminosity plane.
This discrepancy shows that a further refinement of the theoretical models and isochrones is greatly needed.

As the main result of our study, all the empirical extinction laws, derived in the fitting of the data by PARSEC, MIST, 
enhanced DSEP, and IAC-BaSTI, agree with each other and with the estimates $A_\mathrm{V}=0.12\pm0.02$, 
$E(B-V)=0.04\pm0.01$ mag.
These estimates perfectly agree with the recent one for NGC\,6205 from \citet{wagner-kaiser2017}, being twice as high as generally accepted.
Our findings are in line with the results of \citet{gm2017,gm2017big,gm2018}, who have shown that the extinction 
and reddening at high Galactic latitudes behind the Galactic dust layer (the case of NGC\,6205) has been underestimated 
by the 2D maps of \citetalias{sfd} and \citet{2015ApJ...798...88M}, as well as by some 3D maps and models, 
which tend to provide $E(B-V)\approx0.02$ mag for NGC\,6205.

The most probable distance is $7.4\pm0.2$ kpc.
Hence, we found true distance modulus $(m-M)_0=14.35\pm0.06$ and apparent $V$-band distance modulus $(m-M)_\mathrm{V}=14.47\pm0.07$.
These estimates agree with the literature estimates.
The obtained parallax is $0.135\pm0.004$ mas, which agrees within the uncertainties with the recent estimate of the 
parallax from {\it Gaia} DR2 $0.109\pm0.04$ mas.

The age estimates vary even by use of only the optical bands: from $12.3\pm0.7$ Gyr for He--$\alpha$--enhanced DSEP 
to $14.4\pm0.7$ Gyr for MIST.
The most probable age is $13.5\pm1$ Gyr.

Our study suggests a fruitful investigation of other Galactic GCs using the same approach.

\section*{Acknowledgements}

We thank the anonymous reviewer for useful comments.
We thank Santi Cassisi for providing the valuable BaSTI isochrones and his useful comments.
We thank Charles Bonatto for discussion of differential reddening.
We thank Frank Grundahl and Alessandro Savino for providing the Str\"omgren photometric results with their discussion.
This research makes use of Filtergraph \citep{filtergraph}, an online data visualization tool developed at Vanderbilt University through 
the Vanderbilt Initiative in Data-intensive Astrophysics (VIDA) and the Frist Center for Autism and Innovation 
(FCAI, \url{https://filtergraph.com}).
The resources of the Centre de Donn\'ees astronomiques de Strasbourg, Strasbourg, France
(\url{http://cds.u-strasbg.fr}), including the SIMBAD database and the X-Match service, were widely used in this study.
This work has made use of BaSTI, PARSEC, MIST and DSEP web tools.
This work has made use of data from the European Space Agency (ESA) mission {\it Gaia}
(\url{https://www.cosmos.esa.int/gaia}), processed by the {\it Gaia} Data Processing and Analysis Consortium
(DPAC, \url{https://www.cosmos.esa.int/web/gaia/dpac/consortium}).
This study is based on observations made with the NASA/ESA {\it Hubble Space Telescope}.
This publication makes use of data products from the {\it Wide-field Infrared Survey Explorer}, which is a joint project
of the University of California, Los Angeles, and the Jet Propulsion Laboratory/California Institute of Technology.
This work has made use of the Sloan Digital Sky Survey.
This publication makes use of data products from the Pan-STARRS1 Surveys (PS1).

\appendix

\section{Uncertainties}
\label{uncertainties}

The uncertainties of the derived distance, age and reddening depend on the uncertainties of the relative positioning 
of the fiducial sequence and isochrone.

In addition, the uncertainty of the derived reddening includes an uncertainty of the extinction zero-point used.
In turn, it is defined by an uncertainty of the extinction law used for the extinction in the reddest band of 
each dataset.
This provides an uncertainty of the extinction zero-point at a level of several thousandths of a magnitude, 
as discussed in Sect.~\ref{extlaw}.

The most important for the fiducial sequence positioning are:
\begin{itemize}
\item the photometric error from Table~\ref{filters} in combination with the distribution of the stars in the CMD, 
\item the multiple population bias and
\item the differential reddening in combination with other cluster field errors.
\end{itemize}
The most important for the isochrone positioning are:
\begin{itemize}
\item the uncertainty due to a wrong adopted metallicity,
\item the uncertainty due to a wrong adopted He and $\alpha$--enhancement, and
\item the intrinsic systematic uncertainty of the model/isochrone.
\end{itemize}

The latter can be separated into an uncertainty of each model in its application to a dataset
(it seems to be reduced to a level of $<0.01$ mag by our adjustment in Sect.~\ref{adjustment}) and a residual uncertainty of each model itself.
This residual uncertainty can be estimated from the model differences in Fig.~\ref{teff}, if we assume a similar accuracy of the models.

The following quantities are most important to derive distance, age and reddening:
\begin{enumerate}
\item the mean isochrone--fiducial colour difference along the reddening vector (which is rather close to the abscissa 
in the CMDs) for the reddening,
\item the magnitudes of the HB and SGB for the distance,
\item the length of the SGB (the colour difference between the TO and the base of the RGB) and/or the HB--SGB magnitude 
difference -- for the age.
\end{enumerate}

For (i), to determine the fiducial sequence colour, we typically use $10-30$ mag bins. 
For each of them the fiducial colour is determined with a precision of $0.01-0.02$ mag.
Consequently, the uncertainty of the mean is between $0.01/(30)^{0.5}=0.002$ and $0.02/(10)^{0.5}=0.006$ mag.
This is the photometric error contribution to the uncertainty of the fiducial sequence colour.

The remaining uncertainties for (i) are the following.
The uncertainty due to the multiple populations has been discussed in Sect.~\ref{photo}: typically it is 0.006 mag and up to 0.010 mag.
The uncertainty due to the differential reddening and other field errors after the cleaning of the datasets and 
averaging of the errors over the cluster field has been discussed in Sect.~\ref{cleaning}: typically it is 0.005 mag.
We estimate the uncertainties due to wrong adopted metallicity and He--$\alpha$--enhancement empirically, 
i.e. by comparison of the positions of the isochrones with different metallicity and He--$\alpha$--enhancement in the same CMDs.
For the former we vary [Fe/H]$=-1.58$ within $\pm0.04$ stated by \citet{carretta2009}, 
while for the latter we compare the solar-scaled and enhanced isochrones from DSEP or BaSTI.
It appears that each of these uncertainties varies from CMD to CMD between 0.002 and 0.012 mag, 
with a typical value of 0.005 mag and the maximal value for the $U-B$ versus $B$ CMD.
The residual model uncertainty is at a level of $\Delta T_\mathrm{eff}=\pm50$ K, i.e. $\Delta(B-V)=\pm0.015$ mag,
as seen from Fig.~\ref{teff} and discussed in Sect.~\ref{adjustment}.

Adding all the uncertainties, we estimate the total predicted uncertainty of the derived reddening.
It is presented for each CMD in the last column of Table~\ref{fit}.
For almost all CMDs this uncertainty is between 0.03 and 0.04 mag.

For (ii), to determine the magnitudes of the fiducial sequence's HB and SGB, we use, typically, several (e.g. six) colour bins. 
For each of them the magnitude is determined with a precision of a half of one magnitude bin, i.e. $0.1-0.2$ mag.
Hence, the uncertainty of the mean magnitude of the fiducial sequence is between $0.1/(6)^{0.5}=0.04$ and $0.2/(6)^{0.5}=0.08$ mag, 
or slightly better, if we combine the independent estimates for the SGB and HB.
Figs.~\ref{f275_336}--\ref{bv_dr} show that the uncertainties due to the multiple populations and field errors have little effect on the 
fiducial sequence's HB and SGB.
Thus, adding some insignificant multiple population bias and field error, we estimate a typical distance modulus uncertainty,
due to the uncertainty of the magnitudes of the fiducial sequence's HB and SGB,
as about $\pm0.1$ mag. For NGC\,6205 this corresponds to a typical distance uncertainty $\pm340$ pc.

The remaining uncertainties for (ii) are the following.
Fig.~\ref{teff} and all the CMDs show that the uncertainty of the isochrone's HB and SGB magnitude due to 
a wrong adopted metallicity or enhancement is negligible w.r.t. the residual intrinsic uncertainty of the models.
The latter can be predicted from the scatter of the different isochrones in Fig.~\ref{teff} along $L$.
This scatter at the HB or SGB corresponds to a typical distance uncertainty $\pm430$ or $\pm300$ pc, 
if the outlying enhanced BaSTI isochrone is taken into account or not, respectively.
Hereafter we use the latter value, since the enhanced BaSTI is not used in the calculation of the final estimates of 
the distance and age.
Finally, the uncertainty $\pm340$ pc of the fiducial sequence's HB and SGB magnitudes dominates in the total predicted distance uncertainty.

For (iii), the photometric error contribution is the following.
Typically, the length of the SGB is determined with a precision of the colours of few bins engaged at 
the TO and base of the RGB, i.e. $0.01-0.02$ mag, and, hence, with a slightly better uncertainty of the mean.
This corresponds to an age uncertainty within $0.5-1.0$ Gyr.
For the HB--SGB magnitude difference we use several colour bins at the HB and SGB and, as for the case (ii), 
the uncertainty of the mean is typically $0.04-0.08$ mag.
This corresponds to an age uncertainty within $0.4-0.8$ Gyr.
The contribution of the multiple populations into the age uncertainty is discussed in Sect.~\ref{photo}: it is within $\pm0.5$ Gyr.
Adding some field error, we estimate a typical age uncertainty due to the uncertainty of the fiducial sequence, as about $\pm1$ Gyr.

For (iii), similarly to (ii), Fig.~\ref{teff} and all the CMDs show that the uncertainty of the derived age due to the uncertainty of the 
isochrone is dominated by the residual intrinsic uncertainty of the models.
A typical uncertainty of the model HB--SGB magnitude difference and the length of the SGB can roughly be predicted from 
Fig.~\ref{teff} as corresponding to an age uncertainty about $\pm1$ Gyr.
However, an additional uncertainty of the SGB length may be added due to some uncertainties of the colour--$T_\mathrm{eff}$ 
relations and bolometric corrections used, which are not taken into account in Fig.~\ref{teff}.

In some cases the length of the SGB and the HB--SGB magnitude difference provide contradictory results.
For example, in Fig.~\ref{bv}~(c) the IAC-BaSTI isochrone of 14.5 Gyr is a compromise between too long an SGB and too 
big an HB--SGB magnitude difference: 
a shorter observed SGB better fits an age about 15 Gyr, while a shorter observed HB--SGB difference better fits an 
age about 14 Gyr.
In such a case, our search for a minimal total offset between the isochrone and fiducial points provides a 
compromise solution.

Finally, this analysis of the predicted uncertainties shows that the intrinsic systematic uncertainty of the models and isochrones
appears to be one of the main contributions into the total predicted uncertainty.
This shows a need for a further refinement of the models and isochrones.

\section{Some CMDs of NGC\,6205}
\label{cmds}

\begin{figure}
\includegraphics{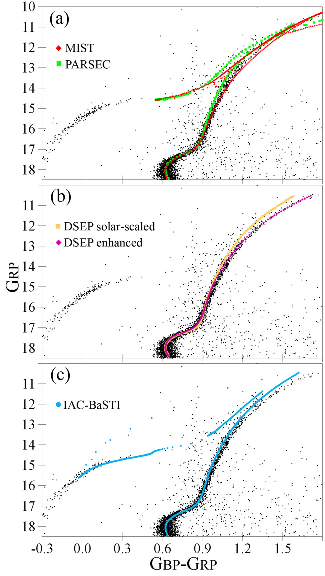}
\caption{The same as Fig.~\ref{bv} but for {\it Gaia} DR2 $G_\mathrm{BP}-G_\mathrm{RP}$ versus $G_\mathrm{RP}$.
}
\label{gbpgrp}
\end{figure}

\begin{figure}
\includegraphics{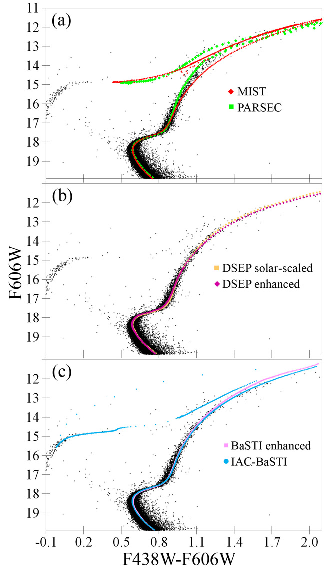}
\caption{The same as Fig.~\ref{bv} but for {\it HST} ACS/WFC3 $F438W-F606W$ versus $F606W$ based on the data 
from \citet{nardiello2018}.
}
\label{f438_f606}
\end{figure}

\begin{figure}
\includegraphics{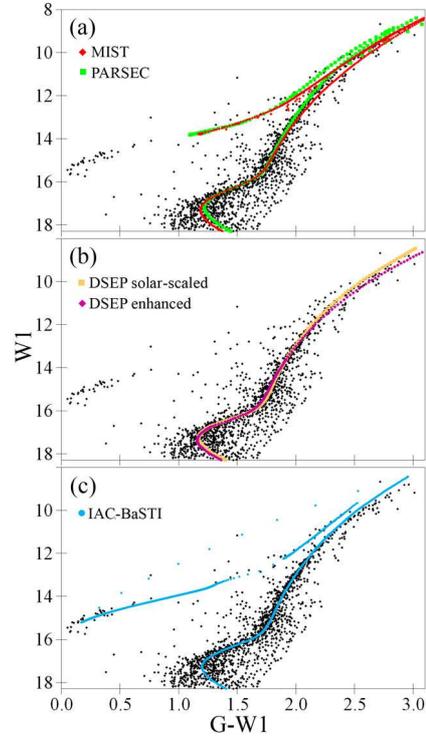}
\caption{The same as Fig.~\ref{bv} but for $G-W1$ versus $W1$ based on the data from {\it Gaia} DR2 and unWISE.
}
\label{gw1}
\end{figure}

\begin{figure}
\includegraphics{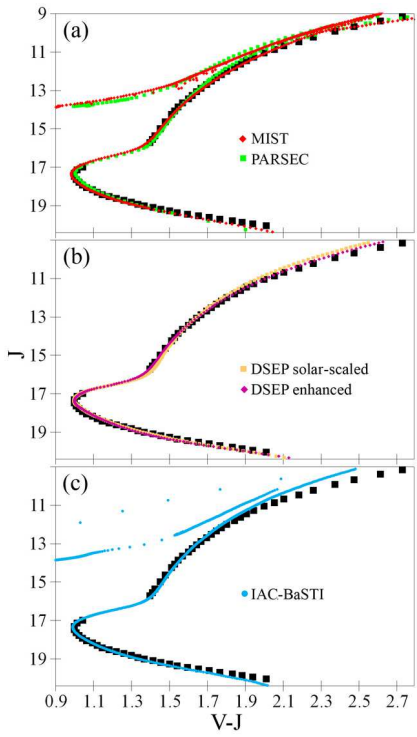}
\caption{$V-J$ versus $J$ CMD of NGC\,6205 with the fiducial sequence from \citetalias{brasseur2010} (black squares), 
and the isochrones calculated with the best-fitting parameters from Table~\ref{fit} and coloured as in Fig.~\ref{bv}.
}
\label{vj}
\end{figure}

\bsp	
\label{lastpage}
\end{document}